%% file: HIGroups_main.tex
\documentclass[fleqn,usenatbib]{mnras}
\usepackage[british]{babel}             
\usepackage{newtxtext}                  
\usepackage{newtxmath}                   
\let\la=\lesssim 
\let\ga=\gtrsim
\usepackage[T1]{fontenc}                
\usepackage{bm}
\usepackage[pdftex]{graphicx}
\usepackage{mathtools}
\usepackage{times}
\usepackage{url}
\usepackage{soul}
\usepackage{float}
\usepackage{caption}
\usepackage[skip=0.5ex]{subcaption}
\usepackage{lipsum}
\usepackage{xcolor}  
\usepackage{comment}
\usepackage{siunitx}
\usepackage{booktabs}
\usepackage{adjustbox}
\usepackage{amsmath}
\usepackage{makecell} 

\newcommand{\spose}[1]{{\hbox to 0pt{#1\hss}}}
\newcommand{\lta}{\mathrel{\spose{\lower 3pt\hbox{$\mathchar"218$}}
     \raise 2.0pt\hbox{$\mathchar"13C$}}}
\newcommand{\gta}{\mathrel{\spose{\lower 3pt\hbox{$\mathchar"218$}}
     \raise 2.0pt\hbox{$\mathchar"13E$}}}

\newcommand{\HI} {H\,{\sc i}} 

\title [GAMA: Group HI Content]{Galaxy And Mass Assembly
  (GAMA): The group \HI\ mass as a function of halo mass}

\author[Dev et al.]
{Ajay Dev,$^{1}$\thanks{E-mail: ajay.dev@research.uwa.edu.au}
Simon P. Driver,${^1}$
  Martin Meyer,$^{1}$
  Sambit Roychowdhury,$^{1,2,3}$ \and
  Jonghwan Rhee,$^{1,2}$
  Adam R. H. Stevens,$^{1}$
  Claudia del P. Lagos,$^{1,2,4}$  \and
  Joss Bland-Hawthorn,$^{5}$
   Barbara Catinella,$^{1,2}$
  A. M. Hopkins,$^{6}$
  Jonathan Loveday,$^{7}$ \and
  Danail Obreschkow,$^{1,2}$
  Steven Phillipps,$^{8}$
  Aaron S. G. Robotham$^{1}$
 \\\\
{}$^1$International Centre for Radio Astronomy Research, University of Western Australia, M468, 35 Stirling Highway, Perth, WA 6009, Australia.\\
{}$^2$ARC Centre of Excellence for All Sky Astrophysics in 3 Dimensions (ASTRO 3D), Australia.\\
{}$^3$University Observatory Munich (USM), Scheinerstr. 1, 81679 Muenchen, German.\\
{}$^4$Cosmic Dawn Center (DAWN), Denmark.\\
{}$^5$Sydney Institute for Astronomy, School of Physics A28, University of Sydney, NSW 2006, Australia.\\
$^6$Australian Astronomical Optics, Macquarie University, 105 Delhi Rd, North Ryde NSW 2113, Australia.\\
{}$^7$Astronomy Centre, University of Sussex, Falmer, Brighton, BN1 9QH, UK.\\ 
{}$^8$H.H. Wills Physics Laboratory, University of Bristol, Tyndall Avenue, Bristol, BS8 1TL, UK.\\
}

\begin{document}
\date{}
\renewcommand{\refname}{REFERENCES}

\pagerange{\pageref{firstpage}--\pageref{lastpage}} \pubyear{2023}

\maketitle

\label{firstpage}

\begin{abstract}
We determine the atomic hydrogen (\HI) to halo mass relation (HIHM) using Arecibo Legacy Fast ALFA survey \HI\ data at the location of optically selected groups from the Galaxy and Mass Assembly (GAMA) survey. We make direct \HI\ detections for 37 GAMA groups. Using \HI\ group spectral stacking of 345 groups, we study the group \HI\ content as function of halo mass across a halo mass range of $10^{11} - 10^{14.7}\text{ M}_\odot$. We also correct our results for Eddington bias. We find that the group \HI\ mass generally rises as a function of halo mass from $1.3\%$ of the halo mass at $10^{11.6} \text{M}_\odot$ to $0.4\%$ at $10^{13.7} \text{M}_\odot$ with some indication of flattening towards the high-mass end. Despite the differences in optical survey limits, group catalogues, and halo mass estimation methods, our results are consistent with previous group \HI-stacking studies. 
Our results are also consistent with mock observations from {\sc Shark} and IllustrisTNG.

\end{abstract}


\input intro

\input data

\input method

\input results

\input disc_and_conc

\section*{Acknowledgements}

We thank the referee for the thorough reading of our work and the constructive feedbacks. GAMA is a joint European-Australasian project based around a spectroscopic campaign using the Anglo-Australian Telescope. The GAMA input catalogue is based on data taken from the Sloan Digital Sky Survey and the UKIRT Infrared Deep Sky Survey. Complementary imaging of the GAMA regions is being obtained by a number of independent survey programmes including GALEX MIS, VST KiDS, VISTA VIKING, WISE, Herschel-ATLAS, GMRT and ASKAP providing UV to radio coverage. GAMA is funded by the STFC (UK), the ARC (Australia), the AAO, and the participating institutions. The GAMA website is http://www.gama-survey.org/ . Based on observations made with ESO Telescopes at the La Silla Paranal Observatory under programme ID 179.A-2004. Based on observations made with ESO Telescopes at the La Silla Paranal Observatory un- der programme ID 177.A-3016. We gratefully acknowledge DUG Technology for their support and HPC services. \\
This work is based on ALFALFA survey conducted with the Arecibo Observatory. We acknowledge the work of the entire ALFALFA team for providing us access to their data cubes. \\
Part of this research was supported by the Australian Research Council Centre of Excellence for All Sky Astrophysics in 3 Dimensions (ASTRO 3D) through project number CE170100013. \\
ARHS is a grateful recipient of the Jim Buckee Fellowship at ICRAR-UWA. 
CL has received funding from the ASTRO 3D, through project number CE170100013, and the Australian Research Council Discovery Project (DP210101945).
DO is a recipient of an Australian Research Council Future Fellowship (FT190100083) funded by the Australian Government.

\section*{Data Availability}
The data underlying this article will be shared upon reasonable request to the corresponding author, AD.

\bibliographystyle{mnras}
\bibliography{HIGroups_main}

\appendix
\input appendix1

\label{lastpage}
\end{document}

%% file: intro.tex
\section{Introduction}
\label{sec:intro_HI}

The effect of environment on the neutral gas content of galaxies is an important driver of galaxy evolution \citep{Cortese2021}. It has been well established that galaxies in higher density environments, such as large groups and clusters, are deficient in neutral atomic hydrogen, \HI, compared to those found in the field \citep{Haynes1984,Solanes2001,Kilborn2009,Chung2009,Cortese2011,Brown2017}.  A number of potentially important effects drive this trend, such as the gravitational interactions between galaxies (tidal stripping, galaxy harassment; \citealt{Moore1996}) and the interaction of galaxies with the intra-group or intra-cluster medium (ram-pressure stripping, strangulation; \citealt{Gunn1972,Balogh2000}). Understanding the balance between these processes, their combined effect, and their evolution with time, remains a core goal for galaxy evolution studies, given the importance of gas as the fuel for star formation and galactic mass assembly.

A particularly useful global observable to constrain these effects is the scaling, and evolution of the \HI\ content as a function of the halo mass, which is dominated by dark-matter (DM).  Models that trace and predict the (multiphase) cool-gas content of galaxies have undergone significant developments in recent years, both in cosmological semi-analytic models \citep{Obreschkow2009,Lagos2011a,Popping2014,Lu2014,Lagos2018},
and hydrodynamical simulations \citep{Lagos2015,Crain2016,Dave2016,Diemer2018,Stevens2019}. \if However, large variations in model predictions still remain due to uncertainties in the underlying physics.\fi Directly relating \HI\ to halo mass provides a valuable observational end-to-end constraint for these models. In doing so, we link the input DM mass distribution for semi-analytic modelling techniques \if, and the dominant mass component driving galaxy evolution,\fi to the desired simulation output product of \HI\ mass. However, making this observational link presents a number of challenges, including the difficulty of detecting \HI\ beyond the relatively local Universe for evolutionary studies, the increasing low-mass limit to which \HI\ can be detected as a function of redshift, and the ancillary dataset requirements for estimating host halo masses.

Deeper \HI\ surveys along with improved galaxy group catalogues have resulted in empirical studies of the \HI-to-halo mass relation (HIHM) being conducted. \citet{Obuljen2018} constrained the HIHM relation as a power law by studying the  abundance and clustering of Sloan Digital Sky Survey (SDSS, \citealt{York2000}) optically selected \HI\ sources from Arecibo Legacy Fast ALFA survey (ALFALFA, \citealt{Giovanelli2005}). \citet{Li2022} estimated the HIHM relation using the \HI\ mass function constructed using an \HI\ mass estimator calibrated with the Extended Galaxy Evolution Explorer Arecibo Sloan Digital Sky Survey (xGASS, \citealt{Cantinella2018}) sample. One method that has undergone substantial development in recent years, which has improved the measurement of \HI\ at larger distances and lower masses, is \HI\ spectral stacking. 

  The spectral stacking technique couples \HI\ survey data with optical redshifts in the same region of sky to enable the co-addition of \HI\ emission from multiple sources, whether or not they are directly detected in the \HI\ image cubes.  The power of this method was first demonstrated for \HI\ by \cite{Zwaan2000} and \cite{Chengalur2001}.  Several studies have since used \HI\ stacking to calculate the \HI\ cosmic density at redshifts higher than those available from blind \HI\ surveys \citep{Lah2007,Delhaize2013,Rhee2013,Rhee2016,Rhee2018,Bera2019,Hu2019,Chowdhury2020,Hu2020,Chen2021}. \cite{Brown2015} used stacking of \HI\ data from the ALFALFA survey to examine the relationship between gas fraction and star formation properties of galaxies.  ALFALFA has also been used to explore the effect of environment through stacking \citep{Fabello2011,Fabello2012} in combination with SDSS data\if citep{Abazajian2009} \fi, finding evidence of ram-pressure stripping in high density environments for galaxies with stellar mass less than $10^{10.5} \mathrm{M_\odot}$.  Moving into kinematic studies, \citet{Meyer2016}  used \HI\ stacking to recover the Tully-Fisher relation from \HI\ Parkes All Sky Survey data \citep{Meyer2004}. Recently, \cite{Rhee2023} performed \HI\ stacking of galaxies in different halo mass bins to study the \HI\ content for different color, environmental effects, \HI\ gas fraction scaling relation, \HI-stellar mass relation, cosmic \HI\ density as well as the HIHM relation with the Deep Investigation of Neutral Gas Origins (DINGO, \citealt{Meyer2009b, Rhee2023}) early science data. \cite{Roychowdhury2022} performed group spectral stacking using ALFALFA and DINGO pilot survey data to study the variations of \HI\ mass to stellar mass fraction as a function of stellar mass and star formation rate. 

While clearly having a number of advantages for extending the redshift and mass ranges over which \HI\ can be studied, stacking has some drawbacks. One of the most significant in single-dish data, is the effect of confusion, where the desired \HI\ signal of a stacked target galaxy can be confused with nearby sources due to limitations in telescope spatial resolution \citep{Jones2016}.  This contaminates the recovered average signal and biases the stack, necessitating the use of often substantial corrections (e.g. \citealt{Delhaize2013} needed a correction factor of $\times 5$ to account for confusion at $z=0.1$ using the Parkes 64m telescope). This problem grows as a function of redshift as the projected density of sources in comparison to the beam size increases. Another potential issue is spectroscopic incompleteness of the input source catalogue.  For galaxy-by-galaxy \HI\ stacking, and the use of this to recover the dependence of \HI\ on halo mass, this could cause an underestimate of \HI\ group content if gas-rich members are overlooked.

In this work, we look to overcome some of these issues by applying a \HI\ {\it group}-stacking technique, combining ALFALFA \citep{Giovanelli2005} data with optical data from the Galaxy and Mass Assembly (GAMA, \citealt{Driver2011, Liske2015, Baldry2018, Driver2022dr4}) survey to derive the dependency of group \HI\ content on halo mass. The group-stacking technique has been previously used in \citet{Guo2020} to calculate the HIHM relation using a combination of ALFALFA and SDSS data. SDSS survey covers $\sim35\%$ of the sky, and is complete to an apparent r-band magnitude of 17.77mag \citep{Strauss2002}. On the other hand, the GAMA survey data used in this work is $\sim2$ mag deeper but covers a much smaller area. The large overlap between SDSS and ALFALFA makes the final group sample used by \citet{Guo2020} much larger ($\sim25000 \text{ groups}$) than the one used here ($\sim350 \text{ groups}$). The novelty in this work is the use of GAMA group catalogue, which is deeper and has its group halo masses calculated in a more empirical way using dynamical mass. We also compare our results with simulations and mock observations in this work. 

This paper is organised as follows. Section 2 describes the datasets that we are using from the ALFALFA and GAMA survey. In Section 3, we describe our method for extracting the spectra, stacking them and calculating \HI\ mass. The results are presented in Section 4, along with our measurements of the HIHM. In Section 5, we make comparisons with simulations and Section 6 presents our conclusions. Throughout, we assume a Hubble constant of $H_0 = \text{70 km s}^{-1} \text{Mpc}^{-1}$ and a $\Lambda$CDM cosmology with $\mathrm{\Omega_{\rm M} = 0.3}$, $\mathrm{\Omega_{\Lambda} = 0.7 }$ in calculating distances, comoving volumes and luminosities.

%% file: data.tex
\section{Data}

\begin{figure}
\centering
\includegraphics[width=9cm]{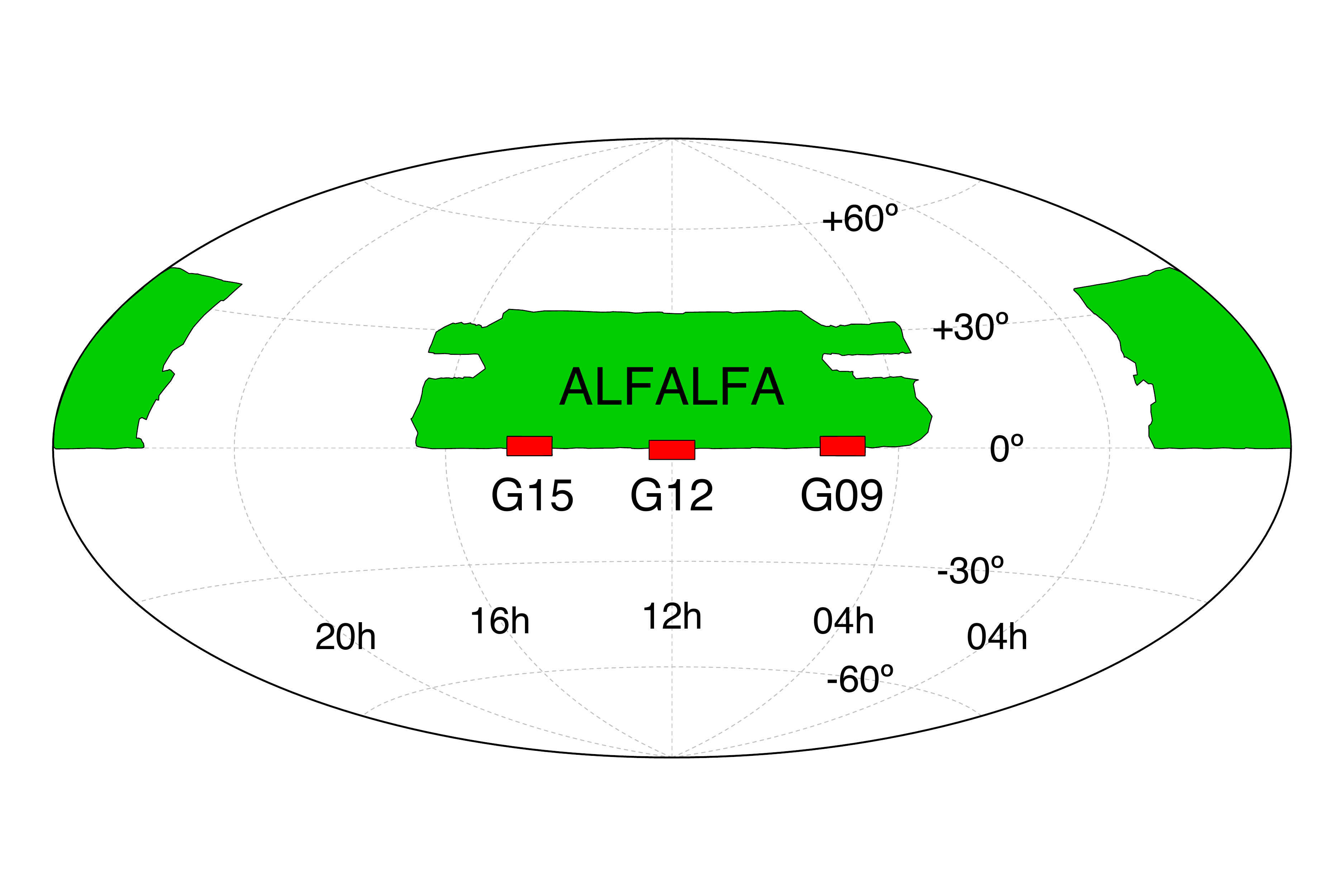}
\caption{On-sky location of the ALFALFA (green) and the GAMA (red) equatorial regions. The overlap includes a sky area of $\sim$ 96 $\text{deg}^2$.}
\label{fig:overlap}
\end{figure}

\subsection{ALFALFA survey}
\label{sec:ALFALFA}
{HI data for this work are taken from the ALFALFA survey, which is a blind HI survey conducted with the single-dish Arecibo telescope using Arecibo L-Band Feed Array (ALFA) \citep{Giovanelli2005}. ALFALFA survey covered a total area of $\sim7000$ $\text{deg}^2$ out to a redshift of $z=0.06$. ALFALFA beams are asymmetric Gaussian with Full Width at Half Maxima (FWHM) of $\ang{;3.8;} \times \ang{;3.3;}$. The ALFALFA grids are $\ang{2.4}$ on each side evenly sampled at $\ang{;1;}$ and the spatial dimensions are $144\times144$ pixels. ALFALFA data have frequency channel spacing of 24.4 kHz with a spectral resolution of 10 $\text{km s}^{-1}$ after smoothing. The average root mean square noise is $\sim2 \text{ mJy channel}^{-1}$. The complete survey footprint can be split into regions with a common declination of $0^\text{o} - 36^\text{o}$ and right ascension of $07^\text{h}30^\text{m} - 16^\text{h}30^\text{m}$ and $22^\text{h} - 3^\text{h}$ (see Fig. \ref{fig:overlap}). \if The survey area partially overlaps with three equatorial GAMA regions giving the opportunity to study the HI content of optically catalogued objects such as galaxy groups. \fi We use the ALFALFA data cubes from the final data release \citep{Haynes2018} for this study.} 

\begin{figure*}
\centering
\includegraphics[width=\textwidth]{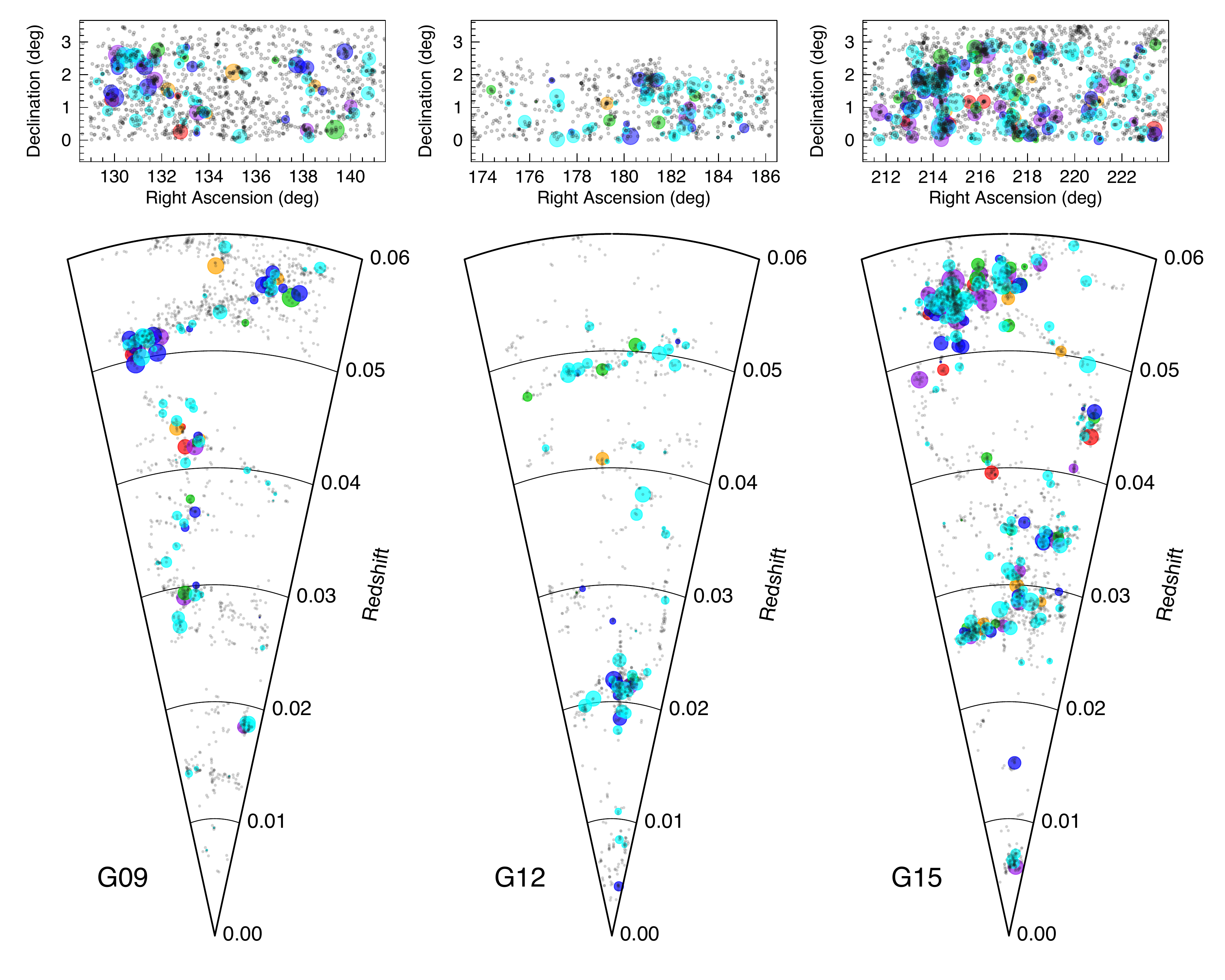}
\caption{Each panel shows a cone plot of the GAMA groups studied in this work. The groups extend to a maximum redshift of $z=0.06$ as shown in the lower panel cones. The upper panel shows the distribution of these groups in Right Ascension and Declination. The groups have been colored according to their multiplicities ($\text{N}_\text{FoF}$ with colors "cyan", "blue", "green", "orange", "red" and "purple" indicating multiplicities of 2, 3, 4, 5, 6 and $\ge$7. The circle size denotes the mass of the groups. The individual GAMA galaxies in these fields are shown as grey points.}
\label{fig:cone}
\end{figure*}

\subsection{GAMA survey}
\label{sec:GAMA}

GAMA is a spectroscopic survey that was conducted using the Anglo-Australian Telescope, in combination with pre-existing literature measurements. It has resulted in over 300,000 spectroscopic redshifts, covering objects up to a magnitude limit of $r < 19.8$, and extending up to a redshift of $z=0.4$. The GAMA survey covered 5 separate fields (G02, G09, G12, G15 \& G23) on the sky with a total area of $\sim$250 $\text{deg}^2$ and spectroscopic completeness of $\sim98.5\%$ in the three equatorial fields where most of the GAMA science is done. It also includes several multi-wavelength photometric and imaging surveys of these regions extending from UV to far-IR. The latest GAMA data release---GAMA DR4 \citep{Driver2022dr4}---provides complete access to the GAMA spectra, redshifts and other data products. One of these data products is the GAMA Group Catalogue which consists of 26,194 galaxy groups created using the friends-of-friends algorithm of \cite{Robotham2011}. The parameters of the algorithm have been extensively calibrated against semi-analytic mock catalogues of the GAMA survey \citep{Merson2012}. The halo masses of the groups are calculated using the velocity dispersion and radius of the group. The halo mass calculation also has been extensively calibrated against the GAMA mock catalogue \citep{Kafle2016}.  We use the latest group catalogue (v10 GroupFinding DMU) taken from the GAMA DR4 website\footnote{\href{http://www.gama-survey.org/dr4/}{http://www.gama-survey.org/dr4/}} for the selection of the galaxy groups studied in this work.

We use the GAMA regions that overlap with the ALFALFA survey, namely G09, G12 and G15 with declination restricted to $> 0^\circ$. \if The region of overlap between GAMA and ALFALFA with the selected group locations is shown in Fig.1.\fi We also restrict the group selection to a redshift of 0.06, which is the maximum depth of the ALFALFA survey. The overlap region covers an area of $\sim 96 \text{ deg}^2$  encompassing a volume of $\sim 1.6 \times 10^5$ $\text{Mpc}^3$. The ALFALFA and GAMA equatorial regions are shown in Fig. \ref{fig:overlap}.

%% file: method.tex
\section{Methods}
\label{sec:HI}

\begin{figure*}
\centering
\includegraphics[width=\textwidth]{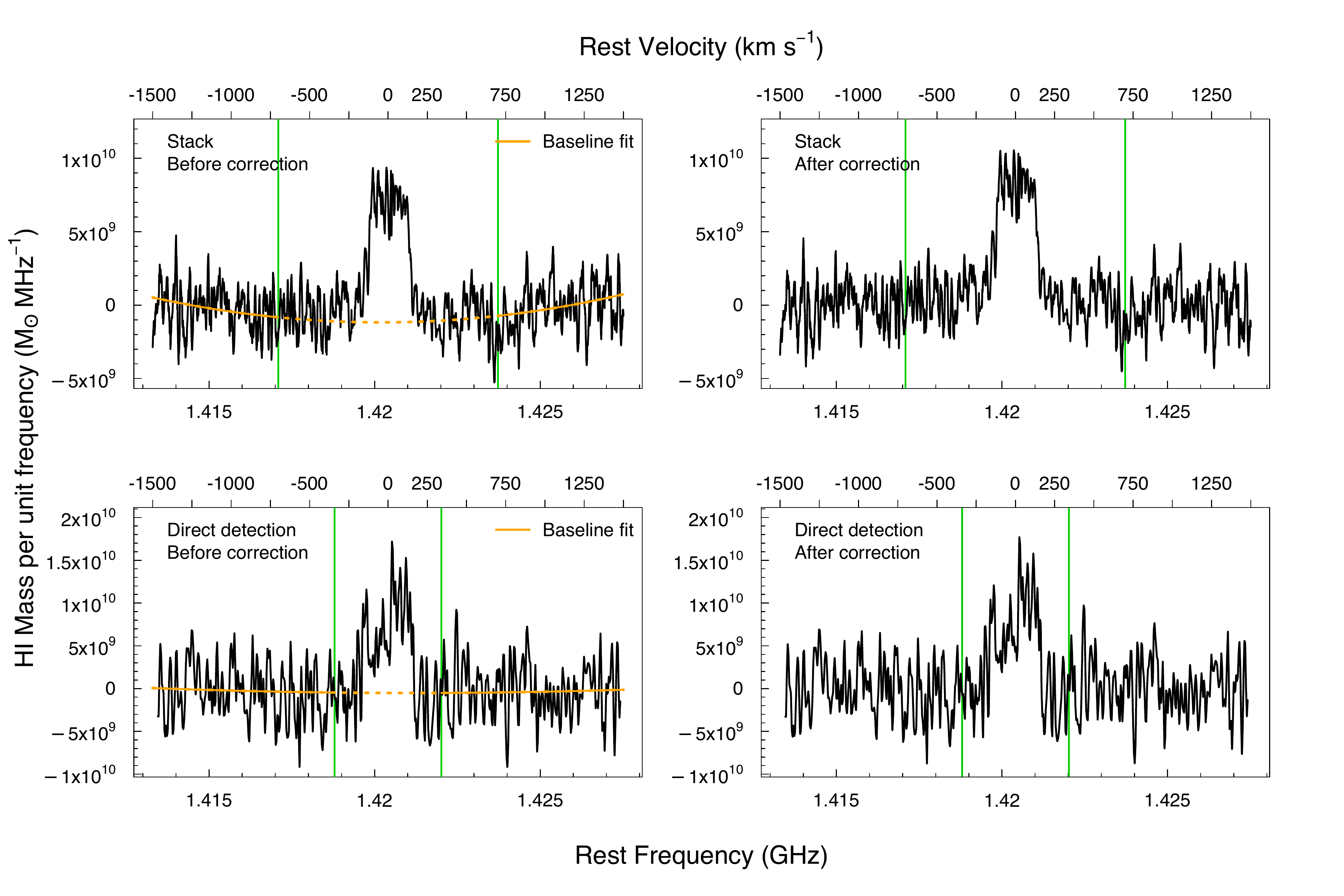}
\caption{Example of second order baseline correction applied to the group spectra. Baseline correction performed on a stacked spectrum and a direct detection is shown on the top and bottom rows respectively. The green line shows the group velocity range for each spectra and the orange line shows the baseline fit done on the data outside the group velocity range. The fit is interpolated for the region in between the group velocity range.  }
\label{fig:baseline}
\end{figure*}

\begin{figure*}
    \centering
    \includegraphics[width=\textwidth]{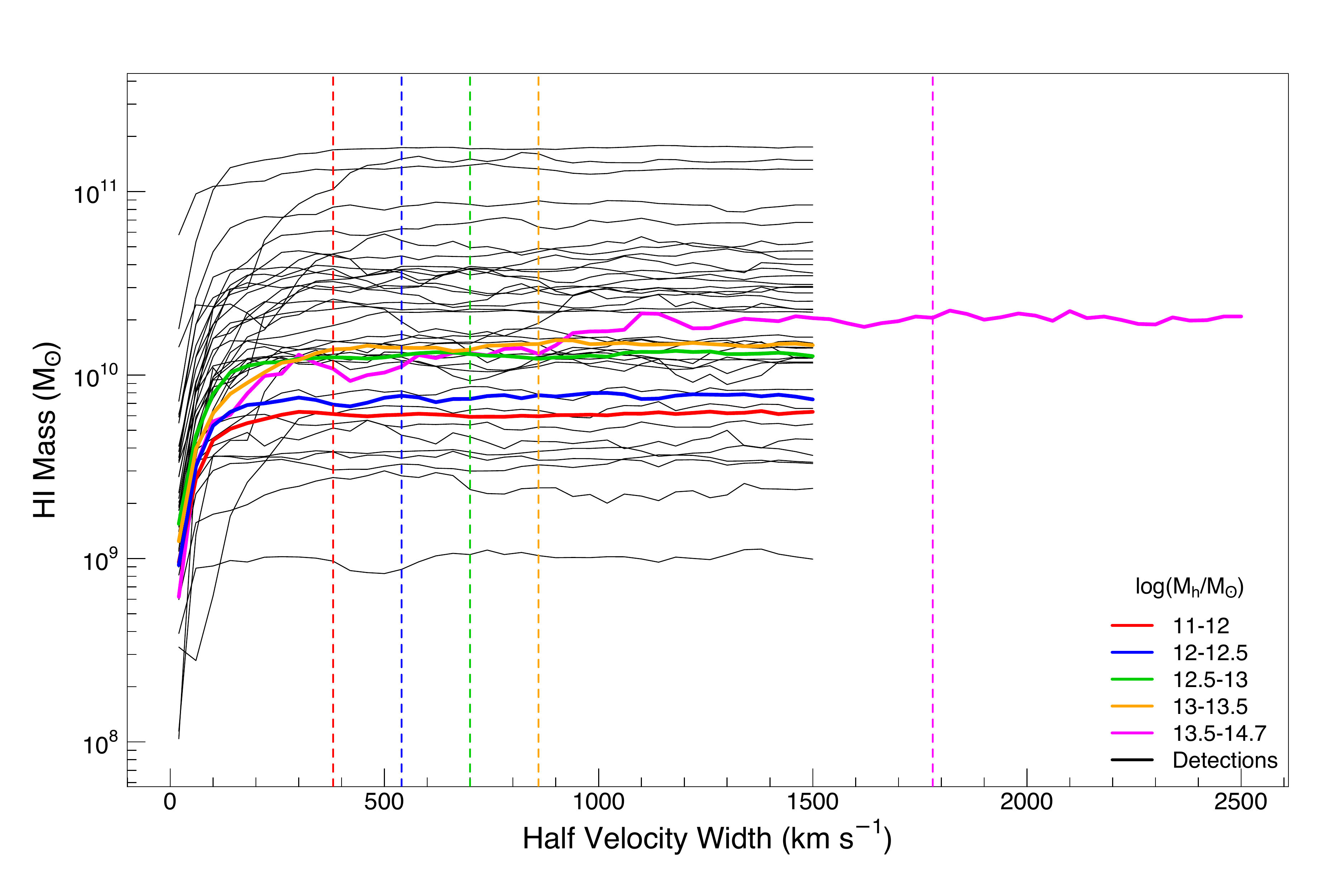}
    \caption{Curve of growth - \HI\ mass as a function of velocity width. In black we show the detections and the stacks of logarithmic halo mass bins 11-12, 12-12.5, 12.5-13, 13-13.5, 13.5-14.7 are shown in red, blue, green, orange and magenta respectively. The vertical colored lines show the group velocity range used for each of the halo mass bin. For all the spectra except the stacked spectrum for the highest halo mass bin, we have spectra extracted over half velocity width of $1500 \text{ km s}^{-1}$. This is the reason all spectra stop at $1500 \text{ km s}^{-1}$ and the spectra for the final mass bin extends up to $ 2500 \text{ km s}^{-1}$.}
    \label{fig:growth_curve}
\end{figure*}

\subsection{Spectra Extraction}
The spectra for the groups are extracted from the ALFALFA data cubes using the group redshift and centre information taken from the GAMA group catalogue. For each group, its \HI\ spectrum is extracted by spatially integrating all the emission inside a  radius given by the quadarture sum of group radius (Rad$_{100}$) and semi-major axis of the beam (\ang{;1.9;}) from the group centre, and using a $\pm 1500 \text{ km s}^{-1}$ rest frame velocity range. $\text{Rad}_{100}$ is defined as the projected distance from the central group member to the most distant group member converted into angular coordinates and \ang{;1.9;} is the semi-major axis of the ALFALFA beam. 

The spectral extraction technique varies depending on whether the object for which the spectrum is being extracted is larger or smaller than the beam size of the telescope. For all our groups, the spectra are extracted with a radius larger than the ALFALFA beam size. So, the extracted spectrum is given by:

\begin{equation}
	S_{\nu} = \frac{1}{C} \times \sum\limits_{x} \sum\limits_{y} S_{\nu}(x,y)  
\end{equation}
where $S_{\nu}(x,y)$ is the flux density of each pixel at position $x$,$y$ and $C$ is a factor correcting from Jy/beam to Jy/pixel (i.e. $C$ = area of beam/area of pixel). The area of the beam is given by:
\begin{equation}
	A_{\mathrm{beam}}=\frac{\uppi b_{\mathrm{min}} b_{\mathrm{max}}}{4 \ln(2)} = 14.21\ \rm{arcmin^2},
\end{equation}
where ${b_\mathrm{min}}$ and ${b_\mathrm{max}}$ are the semi-minor and semi-major axes of the ALFALFA beam. The ALFALFA data are available for two linear polarizations. We perform the spectral extraction for both the polarizations and then take the average of the flux density of both polarizations. For further analysis, we use the polarization averaged spectrum.

\begin{figure*}
\centering
\includegraphics[width=\textwidth]{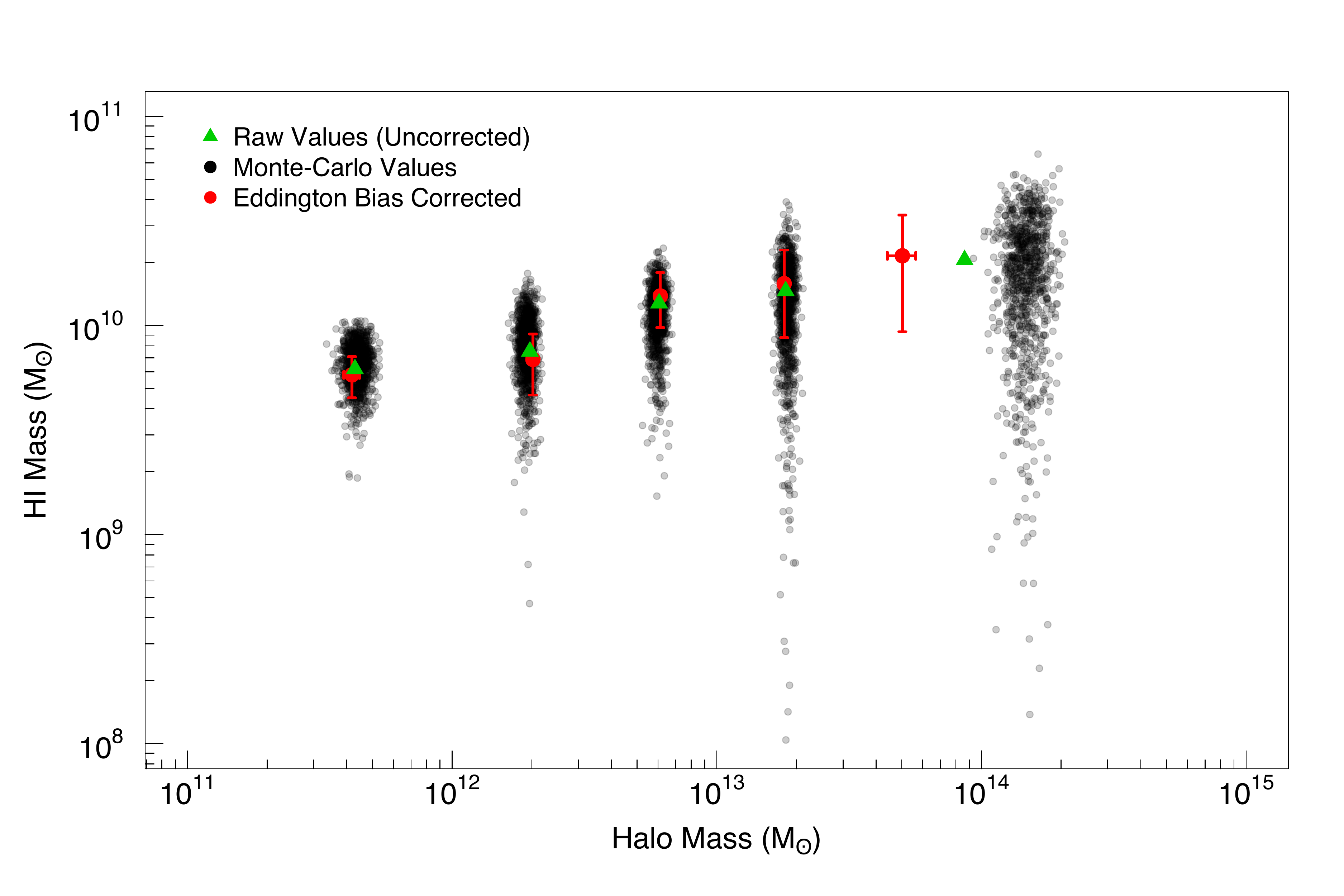}
\caption{Monte Carlo analysis of our group stacking. The raw HIHM values for each halo mass bin is shown in green. In black, the 1000 Monte-Carlo sampled values are shown. The Eddington bias corrected points are shown in red. We can see that significant Eddington bias is only present in the highest mass bin.  }
\label{fig:eddington_bias}
\end{figure*}

\subsection{Group Selection}
{After spectral extraction, we have 501 groups which have overlap between GAMA and ALFALFA. Groups studied in this work have multiplicity of 2 or more ($\textit{N}_\text{FoF} \ge 2$). Halo masses for our groups are taken from the GAMA group catalog (MassAfunc parameter). These are group dynamical masses calculated using group velocity dispersion as given by Eq.18 in \cite{Robotham2011}. We only retain groups which have halo mass listed in the group catalogue. For our study, we only use groups with halo masses above $10^{11} \mathrm{M_\odot}$ as halo masses below this have larger uncertainities. \if The groups which have their \HI\ spectra flagged due to Radio Frequency Interference (RFI) at the region of the rest-frame 21 cm are also taken out from our analysis. \fi We also exclude groups that have more than 50\% of the data flagged due to Radio Frequency Interference (RFI) in the central region where majority of the \HI\ signal is expected. At the end, we have 345 group spectra on which we perform our study. The distribution of these groups on the sky and as a function of redshift is shown in Fig. \ref{fig:cone}. The groups are coloured according to their group multiplicities ($\textit{N}_\text{FoF}$) and their size is representative of their group mass in logarithmic scale. 
}

\subsection{\texorpdfstring{\HI\ Analysis}{HI Analysis}}

\subsubsection{Redshift Correction}

{The extracted spectra are all in the observed frame. So as a first step, we align all the spectra to the rest frequency (1420.4 MHz). This is done by shifting the frequency axis,}
\begin{align}
 && \nu_{\text{rest}}  = \nu_{\text{obs}}\text{(1+\textit{z})}
,\end{align}

\noindent where $\nu_{\text{rest}}$ and $\nu_\text{{obs}}$ are the rest and observed frame frequencies, and \textit{z} is the redshift of the group taken from the group catalogue. We then convert the flux density to \HI\ mass per unit frequency (referred to as the mass spectrum) \citep{Delhaize2013},
\begin{align}
&& \left(\frac{\textit{M}_{\text{\HI},\nu_{\text{obs}}}}{\text{M}_{\odot} \text{MHz}^{-1}}\right) = 4.98\times 10^7 \left(\frac{\textit{S}_{\nu_{\text{obs}}}}{\text{Jy}}\right) \left(\frac{\textit{D}_L}{\text{Mpc}}\right)^2  
,\end{align}

\noindent where $S_{\nu_\text{{obs}}}$ is the observed \HI\ flux density and $D_L$ is the luminosity distance of the group. To conserve the total flux or mass, we also scale the mass spectrum appropriately,
\begin{align}
&& \textit{M}_{\text{\HI},\nu_{\text{rest}}} = \frac{\textit{M}_{\text{\HI},\nu_{\text{obs}}\text{(1+\textit{z})}}}{\text{1+\textit{z}}}. 
\end{align}

\subsubsection{Baseline Correction}

{
Extracted spectra can have a residual baseline offset from 0, so we perform a baseline correction. \citet{Fabello2011} performs a first order baseline correction in their stacking of ALFALFA galaxies. \citet{Guo2020} also mentions that their ALFALFA group stacks were slightly negative and they performed a second-order polynomial baseline correction. In our spectra, to mark the region where there is potential \HI\ signal, we choose the outermost galaxy member of the group in velocity space with respect to the group centre, then add/subtract a 250 $\text{km s}^{-1}$ velocity to it to allow for additional emission spread due to galactic rotation, and mirror this on either side of the spectra centered on the central group member to obtain a symmetric extraction range. This is defined as our group velocity range. We choose 250 $\text{km s}^{-1}$ buffer for the spread due to galaxy rotation as it is similar to the rotational velocity of Milky Way-like galaxies \citep{Eilers2019}. We perform a second order polynomial fit on the part of the spectra outside the group velocity range. After fitting, we interpolate the line within the group region. The total baseline offset calculated this way is then subtracted from the original spectrum to obtain the baseline-corrected spectrum. For the stacked spectrum, we perform the baseline correction after stacking and not on the individual spectrum. Two examples of baseline correction are shown in Fig. \ref{fig:baseline} --- stacked spectrum (top panels) and direct detection spectrum (bottom panels). On the left panels are the spectrum before baseline correction with the orange line showing the second-order baseline correction which will be subtracted from the original spectrum. On the right panels, we show the baseline corrected spectra. The vertical green lines show the group velocity range. 

\subsubsection{\HI\ Mass Determination}

{The \HI\ mass is determined by integrating the \HI\ mass per unit frequency over some channel width,}
\begin{align}
&& \left<\textit{M}_{\text{\HI}}\right> = \int_{\nu_{\text{rest,1}}}^{\nu_{\text{rest,2}}}\textit{M}_{\text{\HI},\nu_{\text{rest}}} \text{d}\nu_{_{\text{rest}}}.
\end{align}

\noindent For the group \HI\ mass, we average the \HI\ mass calculated for increasing widths beyond the group velocity range. We calculate and plot the \HI\ mass for varying group velocity widths, which we call as the curve of growth. This curve is expected to flatten once the \HI\ signal contribution from the group ends and the variations thereafter arise from the noise and background. The curves of growth for all the detections and stacks are shown in Fig. \ref{fig:growth_curve}. We take an average of the \HI\ mass instead of fixing it at the group velocity width in order to consider the variations in the background and to reduce the impact of our choice on the group velocity width. The curve of growth for detections (black lines) and stacks (coloured lines) are shown in Fig. \ref{fig:growth_curve}. 

\begin{figure*}
    \centering
    \includegraphics[width=\textwidth]{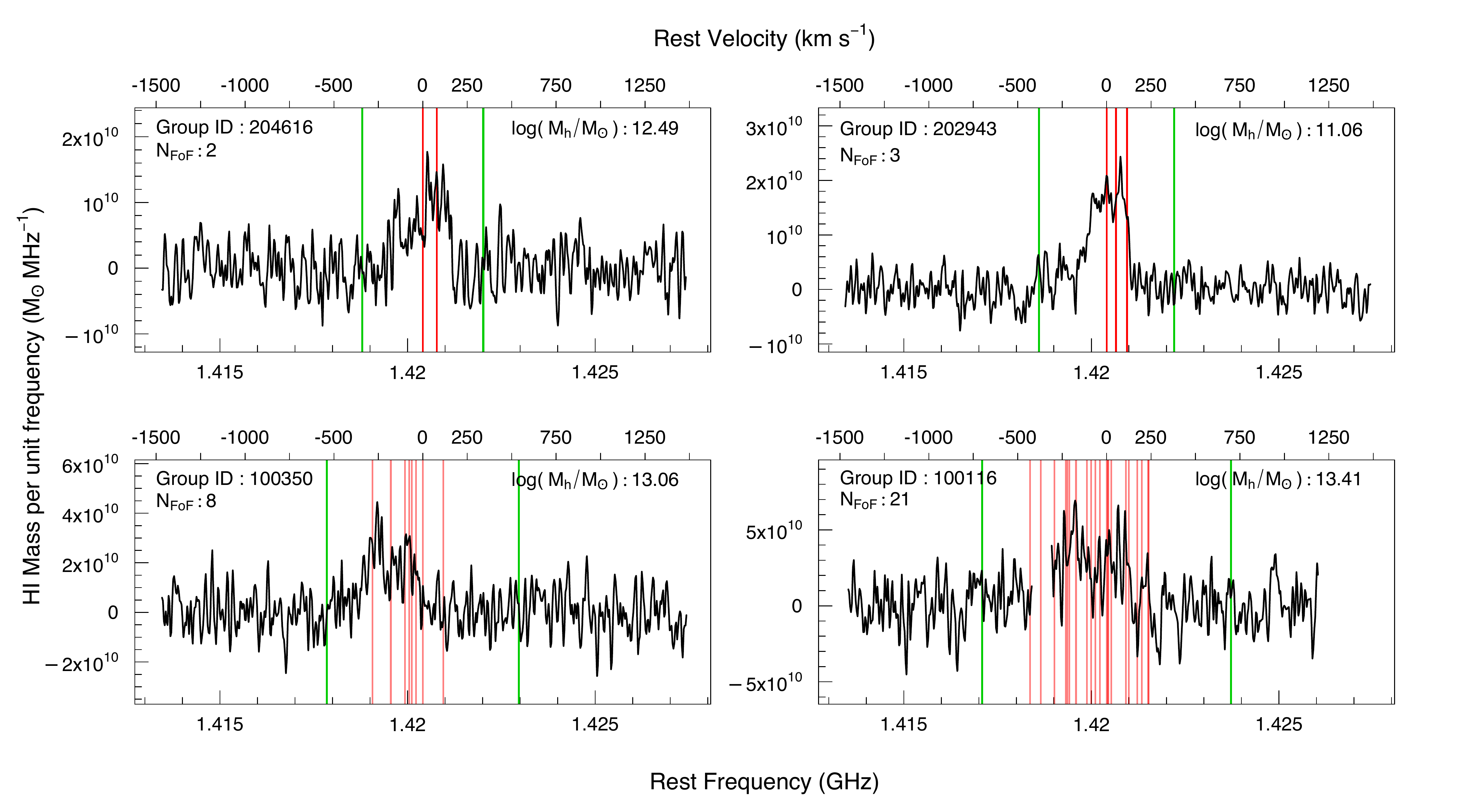}
    \caption{Spectra of 4 (out of 37) groups which have direct detections based on a SNR cut of 3 are shown here. All the spectra have been shifted to the rest frame.  The GAMA Group ID, number of group members and halo mass of the group is shown in each panel. The positions of various group members are shown in red. We show the all the 37 direct detection spectra in Fig. \ref{fig:det1}. }
    \label{fig:detections}
\end{figure*}

\subsubsection{Direct Detection}
Out of the 345 group, we make a signal to noise ratio (SNR) cut of 3 to classify a group spectrum as being a direct detection. The SNR analysis is performed on the observed flux data using the following equation from \citet{Meyer2017}:

\begin{align}
&& \textit{SNR} = \frac{S}{\sigma_{S_{\nu_\text{chan}}}
\Delta \nu_\text{chan} \sqrt{N_\text{chan}}}
\end{align}

\noindent where, \textit{S} is the integrated flux calculated between the group velocity width, $\sigma_{S_{\nu_\text{chan}}}$ is the root mean square flux density error calculated on the spectra excluding the region in between the group velocity width, $\Delta \nu_\text{chan}$ is the channel width and $N_\text{chan}$ is the number of channels over which the flux is measured. \if Any of the group spectra which has $>30\%$ of the data flagged is also excluded as a direct detection even if the SNR is greater than 3.\fi Any of the group spectra that do not have sufficient data to perform baseline correction is also excluded from being classified as a direct detection. We perform baseline correction for these detections and then calculate their \HI\ masses. 
\subsubsection{Stacking}

  Majority of our group \HI\ spectra have low SNR. Hence, we perform stacking to extract the average \HI\ emission. Our stacked spectra includes the contributions from both detections and non-detections. The stacking is done in 5 different halo mass bins --- $\text{log}(\textit{M}_\text{h}/\text{M}_\odot) = 11\text{-}12, 12\text{-}12.5, 12.5\text{-}13, 13\text{-}13.5, >13.5$. These bins have roughly been chosen to have equal number of spectra for stacking ($\sim$75) in each bin. However, in the last bin, we have lower number of spectra to stack (38) due to the relative rareness of high mass groups. Dividing the groups into five halo mass bins enables us to study the properties of groups in a broader range of environments. Further information about the bin size and number of spectra stacked in each bin are shown in Table 1. For stacking, we first shift all the group spectra to their rest frame using Eq. 5. All the aligned mass spectra are then co-added and averaged with equal weighting applied to each of the spectrum. In many other analysis, inverse-variance weighting is often used (e.g. \citealt{Guo2020, Roychowdhury2022, Rhee2023}). In our case with relatively small numbers of groups in each halo mass bin, we prefer the equal weighting method so as not to skew the result overly towards a single dominating group. RFI masked regions in every group spectrum is given zero weighting while performing the stacking. To calculate the velocity range for the stacked spectrum, we choose the most outlier galaxy in each halo mass bin, add/subtract 250 km/s and mirror this on either side of the spectra. We then perform the baseline correction on the stacked spectrum and then calculate the \HI\ mass as mentioned earlier (see Eq. 6) for a single group spectrum. The halo mass of the bin is taken as the mean of the halo mass of all the groups in a particular bin that were involved in the stack. 

\subsection{Error Analysis}

The halo mass errors for the direct detections are calculated using the error corrections in \citet{Robotham2011}. These are based on the number of group members for each group. The halo mass error decreases with increasing group membership from $\sim 1 \text{ dex for \textit{N}}_\text{FoF}=2 \text{ to} \sim 0.1 \text{ dex for \textit{N}}_\text{FoF} \ge 19$. For the errors in \HI\ mass for direct  detections, we include two contributions - the errors arising due to uncertain velocity width and the error due to the noise in the spectrum. First, we look at our \HI\ mass calculation using the curve of growth. Although the curve of growth is supposed to be ideally flat, that is not the case because of background noise. We calculate the 1-sigma variation in the total \HI\ mass by increasing the velocity widths starting from the group velocity range to the end of the spectrum. This is taken as our first error. Next, we try to model the effect of noise in our spectrum. For this, we recreate each detection group spectra by adding a random noise to the flux at each channel which is calculated by sampling from a uniform distribution with mean 0 and standard deviation as the root mean square error of the spectrum. We then recalculate the \HI\ mass for the new spectra. This is repeated 1000 times and the standard deviation from this is taken as the second error. These two errors could be correlated, but we follow the conservative approach and add the errors in quadrature to get the final \HI\ mass error for detections. 

For the stacks, we perform a Monte-Carlo (MC) analysis by recalculating the halo mass for each of the 345 groups from a normal distribution with mean equal to the GAMA group halo mass and standard deviation equal to the halo mass error. This can result in groups moving into different halo mass bins. We now recalculate our full stacking analysis with the new set of halo mass bins and calculate the \HI\ masses and mean of the new halo mass bins. This is repeated 1000 times. The standard deviation from these 1000 iterations is taken as the error for our \HI\ mass and halo mass in each bin.  

We use our MC analysis to quantify and correct for Eddington bias in our data. Eddington bias arises due to the steepness in the Halo Mass Function combined with the large errors in our halo masses tending to scatter more low mass halos towards high than the opposite. We use the Eddington bias correction approach used in \citet{Driver2022hmf}. The results for the Eddington bias correction applied for each mass bin is shown in Fig. \ref{fig:eddington_bias}. Here, we define the difference between the raw value and mean of the MC values as the Eddington bias correction factor. The raw values refer to the \HI\ mass calculated for the original stacked spectra and these are shown by the green triangles. The black circles show the 1000 MC iterations performed by resampling the halo mass of the groups within their errors. Eddington bias correction factor for each halo mass bin is then the difference between the green triangle and mean of the distribution of black points. We subtract this factor once again from the raw values to obtain the original Eddington bias corrected values shown by red dots with errors. The errors are given by the 1$\sigma$ distribution of the black points. The offset between the red, green and black points show the extent of the Eddington bias at each halo mass bin. The first 4 bins do not show a large Eddington bias whereas the bias is significant for the highest mass bin. The large Eddington bias we see in our highest mass bin arises because the “knee” of the halo mass function lies within this bin with a characteristic mass of $\text{log}_{10}(M_\text{h}/\text{M}_\odot)=14.13^{+0.43}_{-0.40}$ \citep{Driver2022hmf}. Hence, far more halos are moved across the knee to high mass than vice versa. We could tighten the upper bound of the bin to mitigate the biases but choose not to due to the already small number of groups in this bin. Note that in these bins we measure the mean halo mass and find that the median has a similar but slightly lower Eddington bias value. The assumption in our Eddington bias correction approach is that the correction factor between first order (difference between red and green point) and second order (difference between green and black points) Eddington bias is the same. We also perform a jack-knife analysis in each halo mass bin to estimate the \HI\ mass error, however we note that this error is subdominant compared to the error from the MC analysis. Hence, for our final stack results, we only use the error from MC analysis.

%% file: results.tex
\section{Results}

\begin{figure*}
\centering
\includegraphics[width=\textwidth]{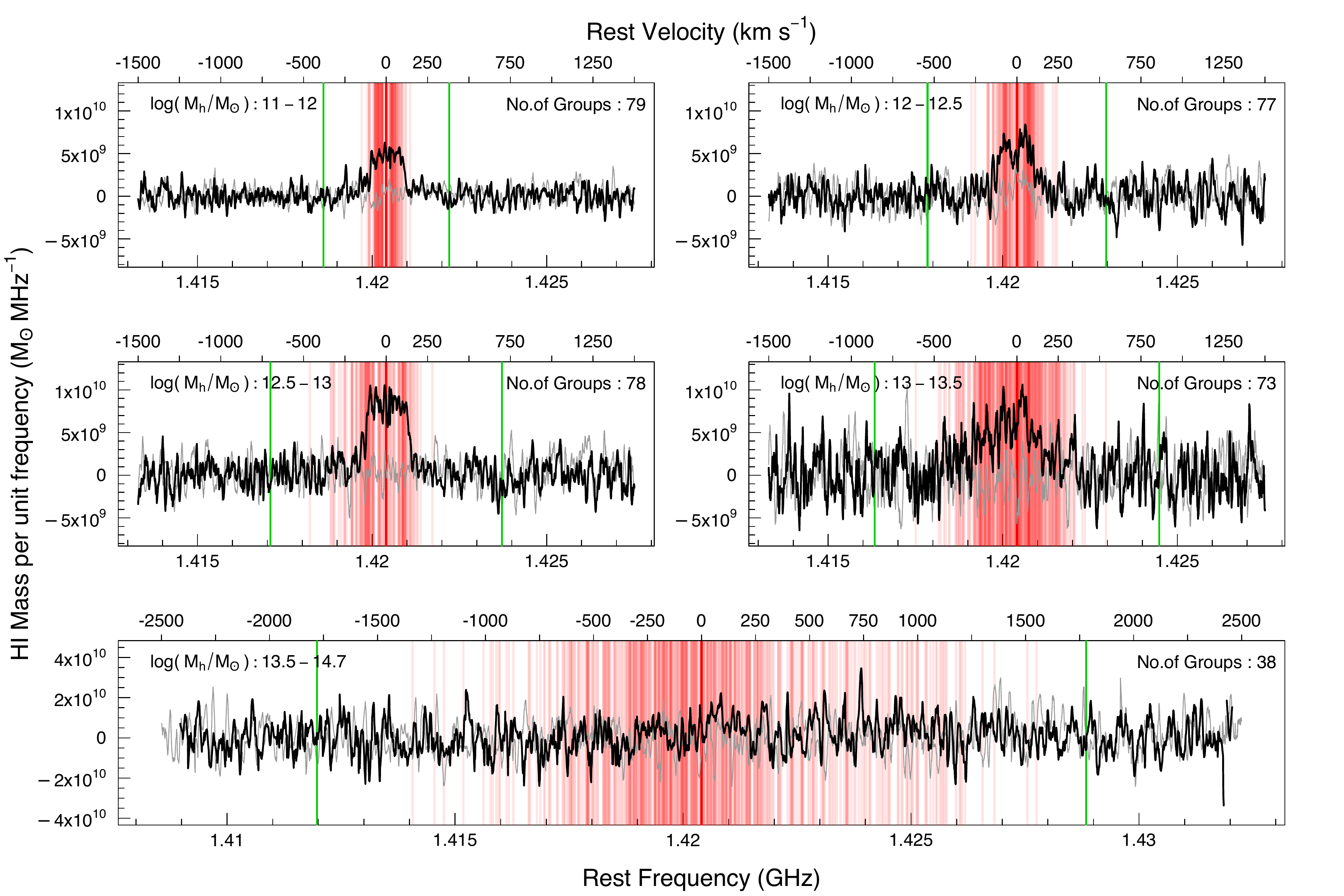}
\caption[]{Stacks for the GAMA groups in halo mass bins. The green line shows the group velocity width of the stacked spectra. On the top side of each panel the halo mass range of the stack and the number of groups used for that stack are mentioned. In red lines, the position of each group member is shown. The groups in the highest halo mass bin have been extracted and stacked over a wider frequency range ($\pm 2500 \text{ km s}^{-1}$) because of its larger group velocity range compared to other 4 bins. In grey, we show a control spectra for each bin where the stacking was performed by randomizing the velocities of each group member.}
\label{fig:stacks}
\end{figure*}

\subsection{Direct Detections}
Based on our SNR cut and group selection, we classify 37 groups out of 345  as being a direct detection. Spectra for 4 of these 37 groups are shown in Fig. \ref{fig:detections}. All the spectra are shown in the rest frame \HI\ velocity space of the central group member. Plotted as vertical red lines are the spectral locations of the group members based on their optical redshifts from GAMA catalogue.  We follow the steps described in Section 3.3 to calculate the \HI\ mass for each of these groups. The curves of growth for all the detections are shown in Fig. \ref{fig:growth_curve} in black.
In Fig. \ref{fig:det1}, we have shown the spectra for all 37 groups having direct detections and the properties of each of these are given in Table \ref{tab:TableA1}.

\begin{figure*}
    \centering
    \includegraphics[width=\textwidth]{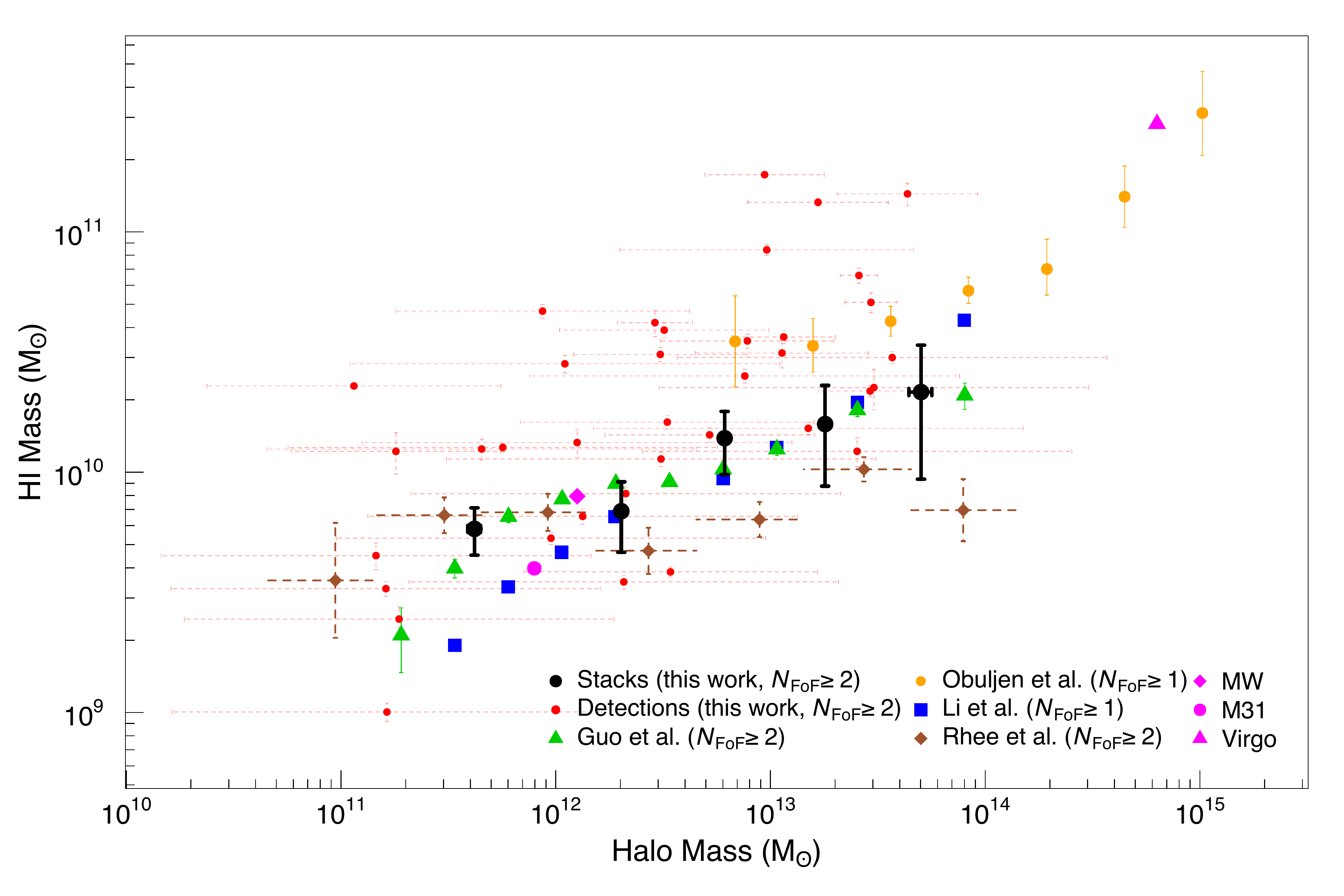}
    \caption[]{\HI-to-Halo Mass Relation for our Group Stacks in halo mass bins as well as individual group detections compared with other observational studies. The stacked groups are shown in black with the error bars calculated using a Monte-carlo Analysis. These points have Eddington bias corrections applied to them. In general, we see an increasing trend in the \HI\ mass as a function of halo mass. The individual group detections are shown as red points. We display the data from \citet{Guo2020}, \citet{Obuljen2018}, \citet{Li2022} and \citet{Rhee2023}. Our method of group \HI\ stacking is most similar to the analysis done by \cite{Guo2020} and we agree well with their trend within the errorbars. \citet{Obuljen2018} \& \citet{Li2022} results are for $N_\text{FoF} \ge 1$ and the rest are for $N_\text{FoF} \ge 2$. We also show the MW, M31 and Virgo measurements from literature.} 
    \label{fig:HIHM}
\end{figure*}

\begingroup
\setlength{\tabcolsep}{10pt} 
\renewcommand{\arraystretch}{1.5} 
\begin{table*}
\begin{tabular}{cccc}
\hline
Halo Mass Bins    & Number of Groups &  $\text{log}_\text{10}(\text{Halo Mass}/\text{M}_\odot)$ & $\text{log}_\text{10}(\text{\HI\ Mass}/\text{M}_\odot)$\\ \hline \hline
$\left[11.0, 12.0 \right)$     & 79                & $11.62{\pm 0.03}$          & $9.76^{+0.10}_{-0.11}$         \\
$\left[12.0, 12.5 \right)$   & 77                & $12.30{\pm 0.02}$          & $9.84^{+0.14}_{-0.19}$         \\ 
$\left[12.5, 13.0 \right)$   & 78                & $12.79{\pm 0.02}$          & $10.14^{+0.14}_{-0.15}$        \\ 
$\left[13.0, 13.5 \right)$   & 73                & $13.25{\pm 0.02}$          & $10.20^{+0.18}_{-0.34}$        \\ 
$\left[13.5, 14.7 \right)$ & 38                & $13.70{\pm 0.05}$          & $10.33^{+0.19}_{-0.37}$         \\ \hline
\end{tabular}
\caption{Group stacking results after Eddington bias Correction. Column 1 - Halo Mass bin, 2 - Number of spectra used in the stack, 3 - Mean Halo Mass of all the groups in the bin expressed in log units, 4 - \HI\ mass of the stacked spectrum expressed in log units. }
\label{tab:Table1}
\end{table*}
\endgroup
\subsection{Stacking Results}

The stacked spectra in 5 different halo mass bins is shown in Fig.~\ref{fig:stacks}. There is a clear signal in all five bins. The halo mass of the groups range from $\textit{M}_\text{h}/\text{M}_{\odot}= 10^{11} \text{ to } 10^{14.7}$. \HI\ mass and halo mass values for all the individual bins are shown in Table \ref{tab:Table1}. The width of the stacked spectrum increases with increasing halo mass, as expected, because the velocity dispersion are higher for groups with more massive haloes. This is also directly visible in the increasing spread of red vertical bands with increasing halo mass bin. The spectrum of the highest mass bin ($= 10^{13.5} \text{-} 10^{14.7} \text{M}_\odot$) is relatively more noisy. This bin also has the least number of spectra stacked compared to the other 4 bins,  mainly due to the diminishing number of high mass groups. For our groups in the highest halo mass bin, we extracted the spectra with a larger velocity range ($\pm 2500$ km/s) compared to groups in other bins ($\pm 1500$ km/s). This is done because the velocity width range for the highest mass bin is more than $\pm 1500\text{km s}^{-1}$ as can be seen in Fig. \ref{fig:stacks}. We also perform a control stack in each of the halo mass bin. For this, we randomize the velocity of each group member and perform the stacking. This should result in a noisy spectra without any signal. The control stacked spectrum for each halo mass bin is shown with the grey line in Fig. \ref{fig:stacks}.

\subsection{The HIHM Relation}
In Fig. \ref{fig:HIHM}, we plot the HIHM relation for our set of groups, both direct detections (red) and stacks (black). The stacked points are shown after Eddington bias corrections. Literature values from \citet{Guo2020}, \citet{Obuljen2018}, \citet{Li2022} and \citet{Rhee2023} are shown. In the HIHM plot, the detections seem to be located above the stacks and this is expected given the detections represent the most \HI\ rich groups for a given halo mass. Hence, the stacks represent the more average \HI\ content for a group in a particular halo mass range by construction. For the stacks, the \HI\ mass increases with halo mass and seems consistent with a power law. Considering the large errorbars, it can also be argued that the HIHM relation seems to flatten towards the high-mass end. We see that the average group \HI\ content is 1.3\% of the halo mass at the lowest mass bin and it comes down to 0.4\% of the halo mass at the highest bin.

Our HIHM results for the stacks is consistent with the results of \citet{Guo2020} across all the halo mass bins. \citet{Guo2020} obtained their HIHM relation using a \HI\ group spectral stacking technique similar to the one used in this work. Their \HI\ data was also from ALFALFA, but the group catalogue is based on SDSS data. The halo masses from the \citet{Lim2017} group catalogue used by \citet{Guo2020} are proxy-masses estimated based on the stellar mass-halo mass relation. This is very different from the mass estimates in GAMA group catalogue \citep{Robotham2011} which corresponds to dynamical halo mass estimates which are calculated using group velocity dispersions. Despite the smaller sample and a different group catalogue, the agreement between our study and \citet{Guo2020} is very good. We have large error bars on our stacked results due to the relatively small group sample used in our stacking along with the implementation of an extensive Monte-Carlo for error analysis. We clarify that these errors are uncertainties on our group-stacking results and do not represent the scatter in HIHM relation.

\citet{Rhee2023} calculated their HIHM relation (shown in brown in Fig. \ref{fig:HIHM}) using the DINGO early-science data and GAMA group catalogue. \HI\ spectra of individual galaxies from groups belonging to common halo mass bins were stacked to calculate the HIHM relation. This is a galaxy-based stacking whereas in our work we do a group-based stacking. Galaxy-based stacking does not take into account the contributions from any \HI\ gas that are unbound to galaxies in the intra-group medium. This would also miss out on the \HI\ in  galaxies that are below the detection limit of GAMA. Hence, the resultant HIHM relation in \citet{Rhee2023} would be considered as a lower limit. It can be argued that our derived group HI masses could be overestimated. The group-stacking technique tries to include all the \HI\ in groups but on the other hand these could also be contaminated by interlopers \citep{Chauhan2021}. 

In Fig. \ref{fig:HIHM}, we show data from \citet{Li2022} in blue. \cite{Li2022} creates a new \HI\ mass estimator where four different galaxy properties---surface stellar mass density, color index, stellar mass and concentration index---are used to model \HI-to-stellar mass ratio. This estimator is calibrated with the xGASS sample \citep{Cantinella2018} and then used to calculate \HI\ mass functions. The final HIHM relation shown in blue is estimated from \citet{Lim2017} group catalog. The \HI\ mass of each group member is calculated using their estimator and summed up to get the total \HI\ mass of a group. All the groups have been divided into halo mass bins to get the final HIHM relation. The \citet{Li2022} results fall below our data at the lowest halo mass bin but predicts more \HI\ in the highest halo mass bin. The disagreement could be attributed to the modified \HI\ mass estimation technique used in \citet{Li2022} compared to the spectral stacking used in this work. It is also important to note that the group membership ($N_\text{FoF}$) in \citet{Li2022} includes $N_\text{FoF} \ge 1$, which can also be the reason for disagreement especially at the lowest mass end because our sample includes groups with $N_\text{FoF} \ge 2$.  

In orange, we have plotted the data from \cite{Obuljen2018}, which uses a similar dataset as the \cite{Guo2020} i.e. ALFALFA and SDSS. However, \citet{Obuljen2018} also does not directly measure the group \HI\ masses but instead it is calculated by integrating the HIMF at various halo masses. Their HIMF mass function is calculated using a 2D stepwise maximum likelihood estimator. \citet{Obuljen2018} probes the high halo mass regime ($\sim 10^{13}\text{-}10^{15} \text{ M}_\odot$). For the halo masses where our study overlaps with \citet{Obuljen2018}, we see that their  \HI\ masses are systematically higher by $\sim 0.3\text{-}0.5$ dex. This could either be because they have systematically overestimated the HIMF for each halo mass, or because of a steep low-mass slope leading to an overestimate while integrating the HIMF. Alternatively, our direct stacking measurements may still be missing very low-mass HI that might have been accounted for in the HIMF. This analysis again includes groups with $N_\text{FoF} \ge 1$ which could bias the comparisons with our results, although the fraction of groups with only one member at the high halo mass end is expected to be low. 

We also show the position of Milky Way (MW), Andromeda (M31) and Virgo cluster in \HI-halo mass plane. The halo masses of MW \citep{Piffl2014} and M31 \citep{Kafle2018} have been calculated by mass modelling based on the escape velocity of stars or planetary nebulae. The Virgo halo mass is taken from \citet{Kashibadze2020}. The \HI\ masses of MW, M31 and Virgo are taken from \citet{Kalberla2009}, \citet{Chemin2009} and \citet{Li2022} respectively.

%% file: disc_and_conc.tex
\section{Comparison with Galaxy formation Simulations}
\label{sec:Simulations}
\begingroup
\renewcommand{\arraystretch}{1.8} 
\begin{table*}
\resizebox{\textwidth}{!}{
\begin{tabular}{ccccccc}
\hline
\textbf{}          & \textbf{Group membership}         & \makecell[cc]{\textbf{Halo mass} \\\textbf{definition}}    & \makecell[cc]{\textbf{Boundary definition}\\ \textbf{for \HI\ mass}}   & \makecell[cc]{\textbf{Includes} \\\textbf{intra-halo \HI\ }} & \textbf{Lightcone} & \textbf{Occupancy} \\ 
\hline\hline

\textbf{\makecell[tc]{Observations \\(this work)}}  & \makecell[tc]{Galaxy FoF \\\citep{Robotham2011}}       & \makecell[tc]{Dynamical \\\citep{Robotham2011}} & \makecell[tc]{$R_\text{2D} = \text{Rad}_{100} + 1.9'$ \\$|\Delta \text{v}| < \text{v}_\text{max}+250 \text{ km s}^{-1}  $}         & Yes                             & N/A                & $N_\text{FoF}>1$   \\

\textbf{\textsc{Shark}-mock 1}             & \makecell[tc]{Galaxy FoF \\\citep{Robotham2011}} & \makecell[tc]{$M_{200c}$ from \\abundance matching}    & \makecell[tc]{$R_\text{2D} < 2R_\text{200c} $\\ |$\Delta$\text{v}| = $\text{v}_\text{vir} - 2\times\text{v}_\text{vir}$}  & No                              & Yes                & $N_\text{FoF}>1$  \\

\textbf{\textsc{Shark}-mock 2}             & \makecell[tc]{Galaxy FoF \\\citep{Robotham2011}} & \makecell[tc]{Dynamical \\\citep{Robotham2011}}  & \makecell[tc]{$R_\text{3D} < R_\text{200c} - 2\times R_\text{200c}$} 
& No                              & Yes                & $N_\text{FoF}>1$  \\

\textbf{\textsc{Shark}-ref}             &   Particle FoF                  &   $M_\text{200c}$ from FoF     & $R_\text{3D} < R_\text{200c}$ 
& No                              & No                & $N_\text{FoF}>0$  \\

\textbf{TNG - mock}             & Particle FoF  & 
\makecell[tc]{Dynamical \\\citep{Robotham2011}}    & All FoF \HI\
& Yes                              & No                & $N_\text{FoF}>1$  \\

\textbf{TNG - ref}             & Particle FoF    &     
\makecell[tc]{$M_\text{vir}$ from FoF \\\citep{BryanNorman1998}}    
& All FoF \HI\
& Yes                              & No                & $N_\text{FoF}>0$  \\ \hline
\end{tabular}}
\caption{Comparison of various simulations and mocks with our observational data. Column 1 - Dataset; Column 2 - indicates how the group membership is defined; Column 3 - halo mass definitions used; Column 4 - indicates the boundary which was used to calculate the \HI\ mass; Column 5 - Indicates whether intra-halo \HI\ would be included in the total \HI\ estimation; Column 6 - Indicates whether mock lightcone was used to for the simulation; Column 7 - Indicates whether the groups include single occupancy halos or not.  }
\label{tab:Table2}
\end{table*}
\endgroup

\begin{figure*}
    \centering
    \includegraphics[width=\textwidth]{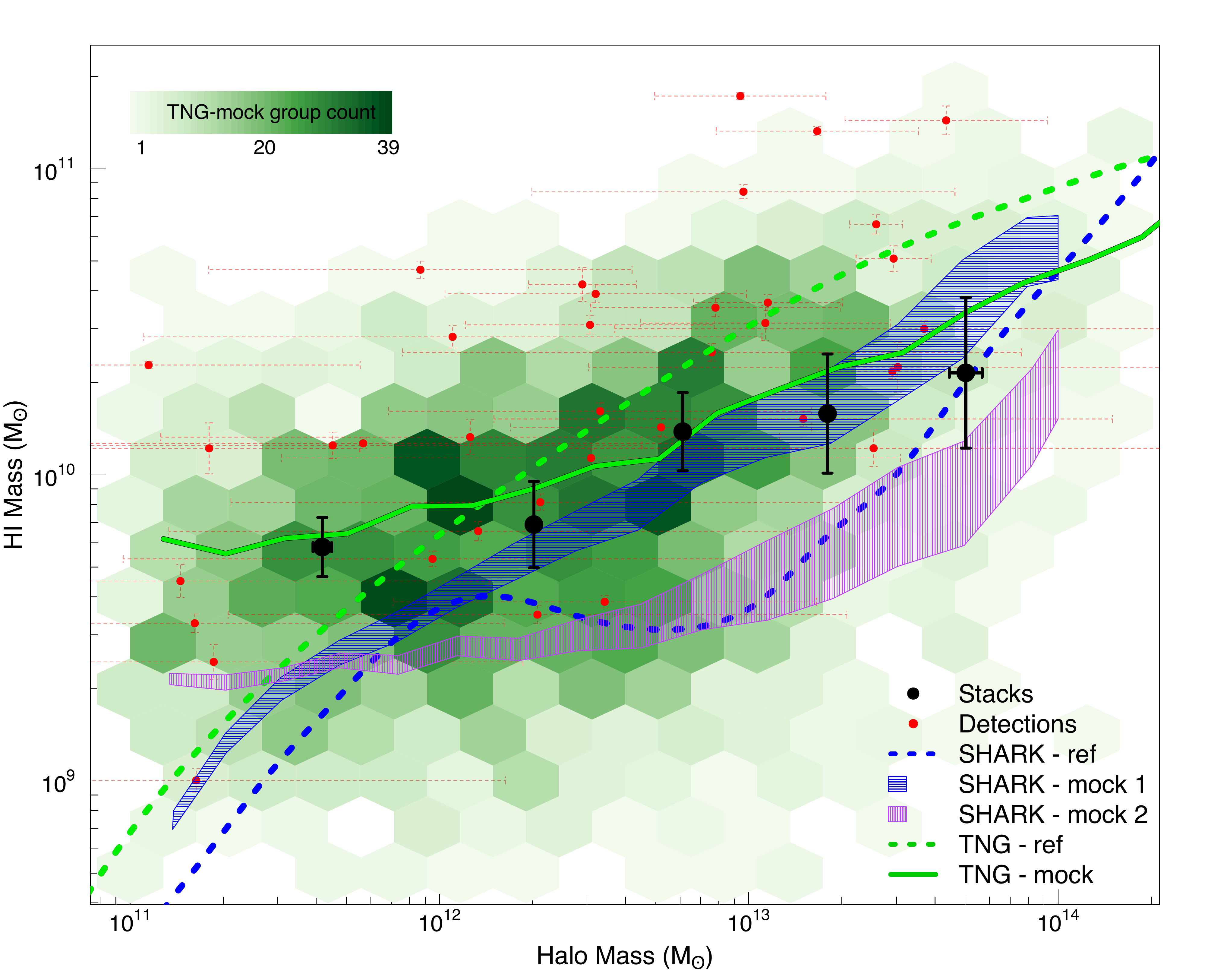}
    \caption{HIHM Relation for our Group Stacks and Direct Detections compared with simulation results from {\sc Shark} (blue) and Illustris TNG (green). The dotted lines represent the raw simulation outputs (\citealt{TNG2018}: TNG-ref; \citealt{Chauhan2020}: {\sc Shark}-ref). The solid lines show mock observations created from the simulations, which are better suited for comparison with our data. Mock {\sc Shark} observations were taken from \citet{Chauhan2021}. The mocks in this case are fully-forward modelled, having simulated a GAMA and ALFALFA sky and having run the \citet{Robotham2011} group finder. {\sc Shark}-mock 1, shown by the blue shaded region, calculates \HI\ mass in a similar manner to our work, but using a larger radius and with a varying velocity window. The halo mass definition, however, is based on abundance matching, and therefore is different to our data. In {\sc Shark}-mock 2, shown by the purple shaded region, the opposite is true; that is, halo mass is defined the same way as in our results, but the \HI\ masses are a direct sum of the known \HI\ masses of the simulated group members. Mock TNG results, shown with green hexagons, were created by selecting FoF groups with 2 or more members, and \HI\ masses were computed with the improved \HI\ / H$_{2}$ modelling developed in \citet{Stevens2019}. The halo masses are dynamical masses, calculated in a similar way as the GAMA group catalogue, but the halo catalogue has not been modified from the public one. The running mean of the TNG mock is shown by the solid green line; this can be fairly compared with our results and shows remarkable agreement. Note that, while TNG-mock is not fully forward modelled as the {\sc Shark}-mock, selection cuts were made to be consistent with GAMA and ALFALFA. \if Our observational results are consistent with mock observations from simulations.\fi We show our 37 direct group detections in red and in green hexagons we show the distribution of the TNG-mock groups in the HIHM plane. The darker hexagons correspond to higher number of groups in a given hexbin. The location of direct detections on the HIHM plot, although naturally biased higher than the stacks as these would be more \HI-rich,  is in agreement with the distribution of TNG groups.}
    \label{fig:sim}
\end{figure*}


We compare the HIHM relation observed by us versus that predicted by the simulations in Fig. \ref{fig:sim}. We use the IllustrisTNG hydrodynamic simulation (\citealt{Nelson2018}; \citealt{Pillepich2018}) and {\sc Shark} semi-analytical model of galaxy formation \citep{Lagos2018}. The HIHM relation has been studied previously using simulations (e.g. \citealt{TNG2018}, \citealt{Spinelli2019, Baugh2019, Chauhan2020}). However, at face value, there is significant variance between different models. The dotted lines in Fig. \ref{fig:sim} show the HIHM relation from the fiducial results of TNG \citep{TNG2018}, referred to as `TNG-ref' hereafter, in green, and {\sc Shark} \citep{Chauhan2020}, referred to as `{\sc Shark}-ref' hereafter, in purple. On the other hand, different observational studies can also differ in their final results. Galaxy group catalogues built from different surveys usually have different magnitude limits, completeness and hence might miss out on the low magnitude galaxies. Halo mass measurement also varies for different group catalogues, e.g. GAMA group catalogue uses a dynamical mass estimation \citep{Robotham2011}, while abundance-matching estimates are used for SDSS-based group catalogues (\citealt{Yang2005, Lim2017}). Therefore, in order to fairly compare the results from observations and simulations, it is necessary to first create mock observations from the same simulations. In Table \ref{tab:Table2}, we provide a comparison between the observation, simulation and mocks in terms of how the groups are defined, the halo mass definition used, the group boundary criteria, whether \HI\ in the intra-halo medium is included, whether lightcones were used to generate the simulations and the minimum number of group members.

We compare our results against mock-observational results from {\sc Shark} \citep{Chauhan2021}, hereafter `{\sc Shark}-mock'. These were created from a GAMA-like mock lightcones, applying an ALFALFA survey selection to it.\citet{Chauhan2021} ran the same \citet{Robotham2011} group finder on the mock lightcone, which allowed them to make a one-to-one comparison with stacking measurements made using FoF-like group finders. \citet{Chauhan2021} extensively looked at various systematics involved in \HI\ stacking measurements such as the effects of different apertures, velocity widths and halo mass estimators. It is important to note here that, mock stacking as referred to in \citet{Chauhan2021} does not involve actual \HI\ spectral extraction of groups and averaging them. All the \HI\ in galaxies within each group as defined by the halo-finder are fully detected and their values are averaged to obtain the \HI\ mass in a given halo mass bin. In Fig. \ref{fig:sim}, as a blue band, we show the HIHM relation from the {\sc Shark}-mock 1, where the group stacking was performed with an adapting group velocity width range that varied depending on the halo virial velocity (adapted from fig. 10 in \citealt{Chauhan2021}). The \HI\ masses were calculated by doing a forward-modelled \HI\ stacking that resembles the realistic geometrical stacking. However, the halo masses in this case were obtained from abundance matching and hence are different from the dynamical halo masses in our stacks. The large scatter between abundance or dynamically estimated masses against the true halo masses in  {\sc Shark} can be seen in fig. 4 of \citet{Chauhan2021}. Dynamical mass estimates would make the relation flatter especially at the low halo mass end. This is shown by the purple shaded region in our Fig. \ref{fig:sim}, {\sc Shark}-mock 2, where the halo masses are dynamical masses. However, the \HI\ mass in this case are not from mock stacking but instead the mean \HI\ content of groups identified by the group finder (adapted from fig. 6 in \citealt{Chauhan2021}). {\sc Shark}-mock 2 groups are defined by a spherical aperture around the central galaxy instead of a projected aperture. The ideal comparison to our results would have been a combination of the two {\sc Shark} mocks with dynamical halo masses and realistic mock stacking to derive the \HI\ content. Excluding the initial point, our measurements are consistent with the {\sc Shark}-mock 1 but have a systematic disagreement with {\sc Shark}-mock 2. The reason for the deviation between {\sc Shark}-mock and {\sc Shark}-ref above $10^{12} \text{M}_\odot$ is also attributed to the large scatter between halo masses estimates in this region, thereby washing out any dip-like feature of the HIHM in {\sc Shark}-mocks that can otherwise be seen in {\sc Shark}-ref \citep{Chauhan2021}.

For the Illustris TNG simulation, we conduct a different procedure that helps us to mimic some of the selection functions we have in the observations, but we caution that we do not build lightcones or run group finders in this case, as was done in \citet{Chauhan2021} for {\sc Shark}; such an analysis is very involved and beyond the scope of this paper. We refer to this as `TNG-mock'. For this, we first select Friends-of-friends (FoF) groups with at least 2 galaxies from the TNG100 simulation (\citealt{Nelson2018}; \citealt{Pillepich2018}) with a stellar mass cut of above $10^{8.5} \text{M}_\odot$.  The galaxies were extracted from a pre-constructed dataset of TNG100 galaxies at $z=0$ with total gas OR stellar masses above 10$^9$\,M$_\odot$ and DM fractions above 5\% (the same dataset used in \citealt{Dave2020}). The selection is not done based on luminosities, which is what would ideally be done with a GAMA-like FoF group finder. The \HI\ masses of the groups were computed using all gas cells in each FoF group. Our calculation of the \HI\ mass of haloes differs to \citet{TNG2018} (shown as green dotted line in Fig. \ref{fig:sim}) in that we have included the updated \HI\ / H$_2$ modelling from \citet{Diemer2018} and \citet{Stevens2019}, using the prescription based on \citet{Gnedin&Draine}. The main differences are the improved modelling of the UV field in each halo, which accounts for local sources from star formation, and the application of a different prescription for how the UV and presence of metals/dust affects the molecular/atomic equilibrium.  We have also excluded haloes occupied by a single galaxy to match the group definition of GAMA. \if Any deviations from the public \HI\ data of TNG100 (see \citealt{Diemer2018}; \citealt{Nelson2019}) have previously been tested to be minimal and should be inconsequential here.\fi To make a fair comparison between our results and TNG100, we recompute the simulation's halo masses to be consistent with the dynamical method of \citet{Robotham2011} used for the GAMA groups. While less ideal, we maintain the FoF group definition as in the public TNG data. The location of these TNG-mock results on the HIHM plot is shown by the green 2D histogram in Fig. \ref{fig:sim} with darker hexbins corresponding to higher number of groups, and we have calculated a running mean shown as solid green line. The mean HIHM relation from the TNG-mock is  consistent with our group stacking results.The reason for the deviation between TNG-mock and TNG-ref is due to the different halo mass definitions. We also plot our direct detections in Fig. \ref{fig:sim} as red circles. The distribution of our direct detections is in agreement with the distribution of groups in the TNG-mock as shown by the 2D histogram. 

Our overall analysis is largely restricted to halo mass below $\sim 10^{14} \text{M}_\odot$ thereby not enabling a proper comparison in the cluster regime. \cite{Obuljen2018} and simulations \citep{Chauhan2020, TNG2018} suggest an increasing trend of the group \HI\ masses even beyond $ 10^{14} \text{M}_\odot$.\if This suggests that cold \HI\ is able to survive ionisation even in such high temperature cluster environments.\fi The better consistency of our results with the mock observations compared to simulations' fiducial results, highlights the importance to create mock catalogues for a fairer comparison between the observations and simulations. The comparisons do show that the way the mocks are observed also affects the resulting HIHM relation \citep{Chauhan2021}. The significant agreement between our data and the predictions of two simulations, which utilized different computational methods, assumptions, and baryon physics implementations, but both based on $\Lambda$CDM, is yet another example of the success of the standard cosmological model.

\section{Conclusions}
\label{sec:conclusion_HI}

In this paper, we have calculated the \HI\ content of galaxy groups in the overlap region of ALFALFA and GAMA surveys. We have used the GAMA group catalogue to determine the location and halo mass of the groups. After spectral extraction and cleaning, we have 345 groups in a halo mass range from $10^{11} \text{M}_\odot$ - $10^{14.7} \text{M}_\odot$ on which we perform our analysis. From the individual group spectra, we have obtained 37 direct detections with SNR $>$ 3. We divide the 345 groups into 5 halo mass bins (see Table \ref{tab:Table1}) and perform the stacking analysis in each bin. We also perform an Eddington bias correction on our stacked results. Eddington bias is prominent only in the highest mass bin.

We look at the HIHM relation for our direct detections and the stacks. The direct group detections are naturally biased to have higher \HI\ masses at a given halo mass than the stacks. But, we see that the location of these direct detections on the HIHM plane is consistent with the mock observations done with TNG. We see an increasing trend of \HI\ mass as a function of halo mass with some indication of flattening towards the higher halo mass end. The average \HI\ mass content changes from $1.3\%$ of the halo mass at $10^{11.62} \text{M}_\odot$ to $0.4\%$ at $10^{13.70} \text{M}_\odot$. We compare our HIHM measurements with previous studies. We find that our results are consistent with the \HI\ group stacking measurements of \citet{Guo2020}. Despite, the differences in group catalogues, survey detection limits as well as the halo mass measurement techniques between our work and \citet{Guo2020}, the results agree across the entire halo mass range. We do find differences with \citet{Li2022} and \citet{Obuljen2018} at the lowest and highest halo mass bins. This could be due to systematic bias in the way the HIHM relation has been calculated in the two studies. We compare our results with simulations and mock observations of {\sc Shark} and TNG. We see that our results are fairly consistent with mock observations and point out that it is essential to create mock observations when making comparisons with theoretical studies. Since, we do not probe the cluster regime in this study, we need future studies of high-mass groups to validate these results. Deeper and more sensitive \HI\ surveys such as DINGO on Australian Square Kilometre Array Pathfinder (ASKAP, \citealt{Johnston2007, Johnston2008, Hotan2021}) and the \HI\ emission project within the MeerKAT International GigaHertz Tiered Extragalactic Exploration  (MIGHTEE-\HI, \citealt{MIGHTEE-1, MIGHTEE-HI}) survey on the Meer-Karoo Array Telescope (MeerKAT, \citealt{MeerKAT}) will enable improved studies of the \HI-to-halo mass relation.

%% file: appendix1.tex
\section{Individual Detections}
We present here the results of the 37 direct detected groups from our study. In Table \ref{tab:Table2}, we provide the properties of all these direct detections, and in Fig. \ref{fig:det1} we show the group spectra of all the direct detections. 

\begingroup
\setlength{\tabcolsep}{10pt} 
\begin{table}
\begin{tabular}{@{}cccc@{}}
\toprule
Group ID & Nfof & $\text{log}_\text{10}(\text{Halo Mass}/\text{M}_\odot)$    & $\text{log}_\text{10}(\text{\HI\ Mass}/\text{M}_\odot)$      \\ \hline \hline
104257   & 2    & 11.21 $\pm$ 1    & 9 $\pm$ 0.04     \\
104903   & 2    & 11.21 $\pm$ 1    & 9.52 $\pm$ 0.03  \\
104942   & 2    & 11.75 $\pm$ 1    & 10.1 $\pm$ 0.02  \\
104959   & 2    & 11.27 $\pm$ 1    & 9.39 $\pm$ 0.06  \\
201275   & 4    & 12.72 $\pm$ 0.49 & 10.16 $\pm$ 0.03 \\
202943   & 3    & 11.06 $\pm$ 0.68 & 10.36 $\pm$ 0.01 \\
202983   & 3    & 12.53 $\pm$ 0.68 & 9.59 $\pm$ 0.02  \\
$206618^*$   & 2    & 12.32 $\pm$ 1    & 9.54 $\pm$ 0.03  \\
300387   & 7    & 12.97 $\pm$ 0.28 & 11.24 $\pm$ 0.01 \\
300394   & 6    & 13.22 $\pm$ 0.33 & 11.12 $\pm$ 0.01 \\
300576   & 5    & 12.89 $\pm$ 0.4  & 10.55 $\pm$ 0.03 \\
300594   & 5    & 12.49 $\pm$ 0.4  & 10.49 $\pm$ 0.03 \\
300618   & 8    & 13.46 $\pm$ 0.24 & 10.34 $\pm$ 0.02 \\
$303979^*$   & 2    & 13.48 $\pm$ 1    & 10.35 $\pm$ 0.08 \\
305114   & 2    & 11.98 $\pm$ 1    & 9.73 $\pm$ 0.03  \\
100116   & 21   & 13.41 $\pm$ 0.09 & 10.82 $\pm$ 0.03 \\
100350   & 8    & 13.06 $\pm$ 0.24 & 10.56 $\pm$ 0.03 \\
$101990^*$   & 3    & 12.98 $\pm$ 0.68 & 10.93 $\pm$ 0.02 \\
104175   & 2    & 11.66 $\pm$ 1    & 10.1 $\pm$ 0.04  \\
105107   & 2    & 13.18 $\pm$ 1    & 10.18 $\pm$ 0.02 \\
200208   & 17   & 13.47 $\pm$ 0.12 & 10.71 $\pm$ 0.04 \\
200556   & 5    & 13.05 $\pm$ 0.4  & 10.5 $\pm$ 0.06  \\
$200857^*$   & 4    & 11.26 $\pm$ 0.49 & 10.09 $\pm$ 0.08 \\
$202650^*$   & 3    & 11.94 $\pm$ 0.68 & 10.67 $\pm$ 0.03 \\
203156   & 2    & 12.88 $\pm$ 1    & 10.4 $\pm$ 0.03  \\
204108   & 2    & 11.16 $\pm$ 1    & 9.65 $\pm$ 0.05  \\
204616   & 2    & 12.49 $\pm$ 1    & 10.06 $\pm$ 0.03 \\
300106   & 11   & 12.46 $\pm$ 0.17 & 10.62 $\pm$ 0.05 \\
301110   & 4    & 12.51 $\pm$ 0.49 & 10.59 $\pm$ 0.03 \\
301438   & 6    & 13.64 $\pm$ 0.33 & 11.16 $\pm$ 0.05 \\
301751   & 3    & 12.52 $\pm$ 0.68 & 10.21 $\pm$ 0.03 \\
303789   & 2    & 12.33 $\pm$ 1    & 9.91 $\pm$ 0.03  \\
303881   & 2    & 12.13 $\pm$ 1    & 9.82 $\pm$ 0.03  \\
304103   & 2    & 12.04 $\pm$ 1    & 10.45 $\pm$ 0.04 \\
305154   & 2    & 13.4 $\pm$ 1     & 10.09 $\pm$ 0.06 \\
305259   & 2    & 13.57 $\pm$ 1    & 10.48 $\pm$ 0.03 \\
305376   & 2    & 12.1 $\pm$ 1     & 10.12 $\pm$ 0.06 \\ \bottomrule
\end{tabular}
\caption{Properties of the direct detections. Column 1 - GAMA Group ID, 2 - Number of group members, 3 -  Halo Mass the group in log units, 4 - \HI\ mass of the group in log units. The Group IDs which have an asterisk (*) are those that have incomplete spectra due to RFI. The \HI\ masses of these groups should be taken as lower limit. }
\label{tab:TableA1}
\end{table}
\endgroup

\begin{figure*}
    \centering
    \includegraphics[width=\textwidth]{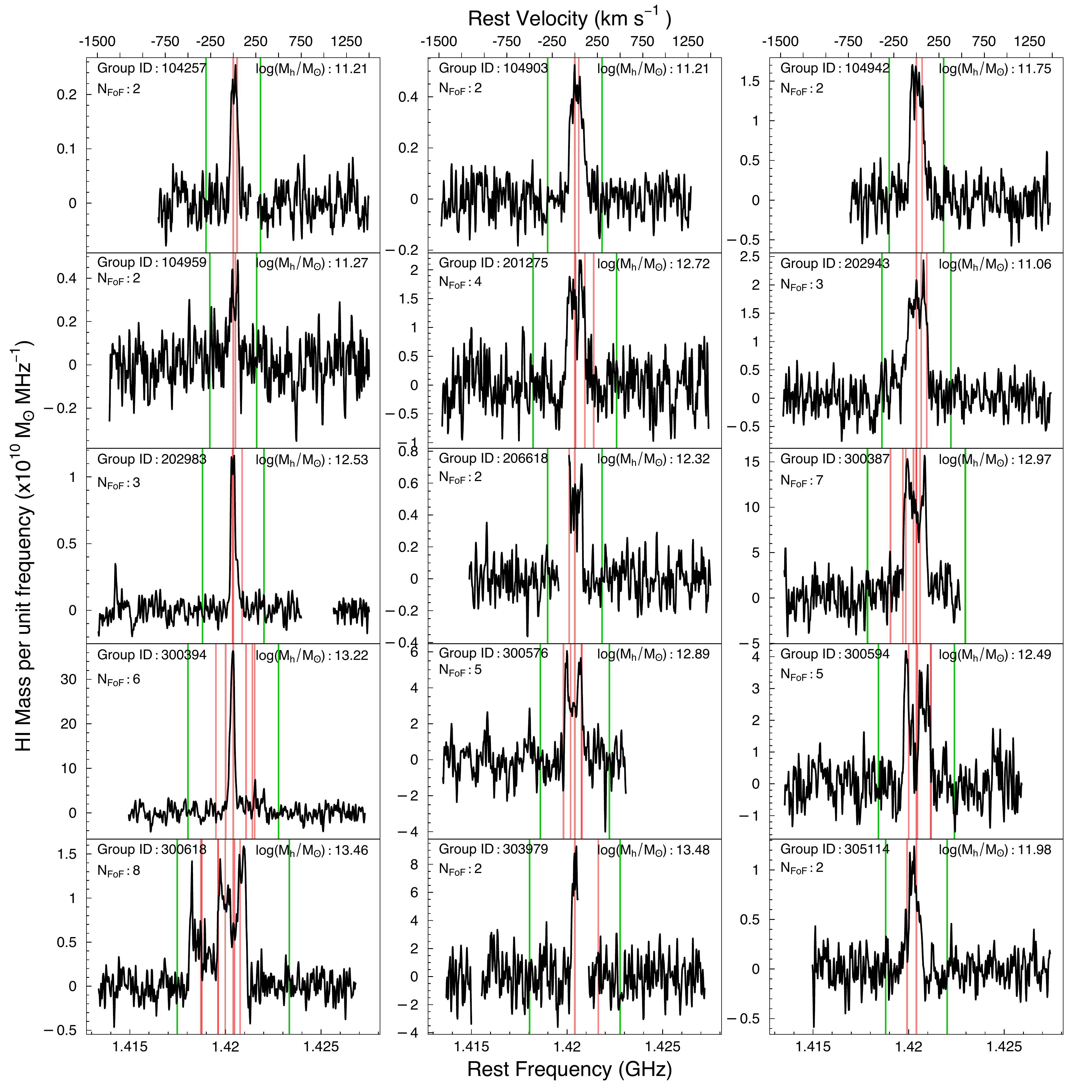}
    \caption{Spectra of all the 37 groups which have detections based on a SNR cut of 3. All the spectra have been shifted to the rest frame.  The GAMA Group ID, number of group members and halo mass of the group is shown in each panel. In red lines, the position of various group members are shown. The vertical green line shows the group velocity range. }
    \label{fig:det1}
\end{figure*}

\begin{figure*}
    \centering
    \includegraphics[width=\textwidth]{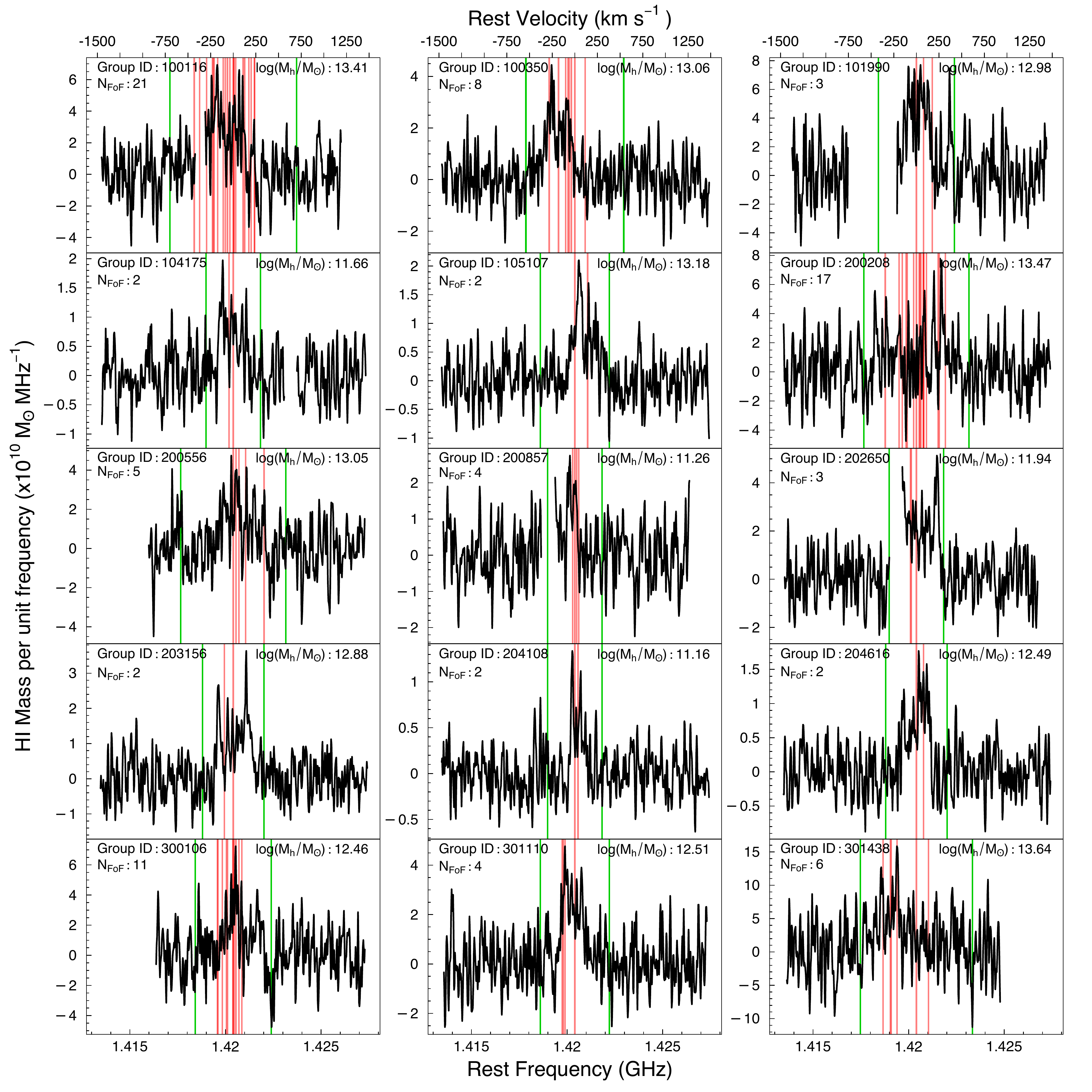}
    \caption*{Continued}
    \label{fig:det2}
\end{figure*}

\begin{figure*}
    \centering
    \includegraphics[width=\textwidth]{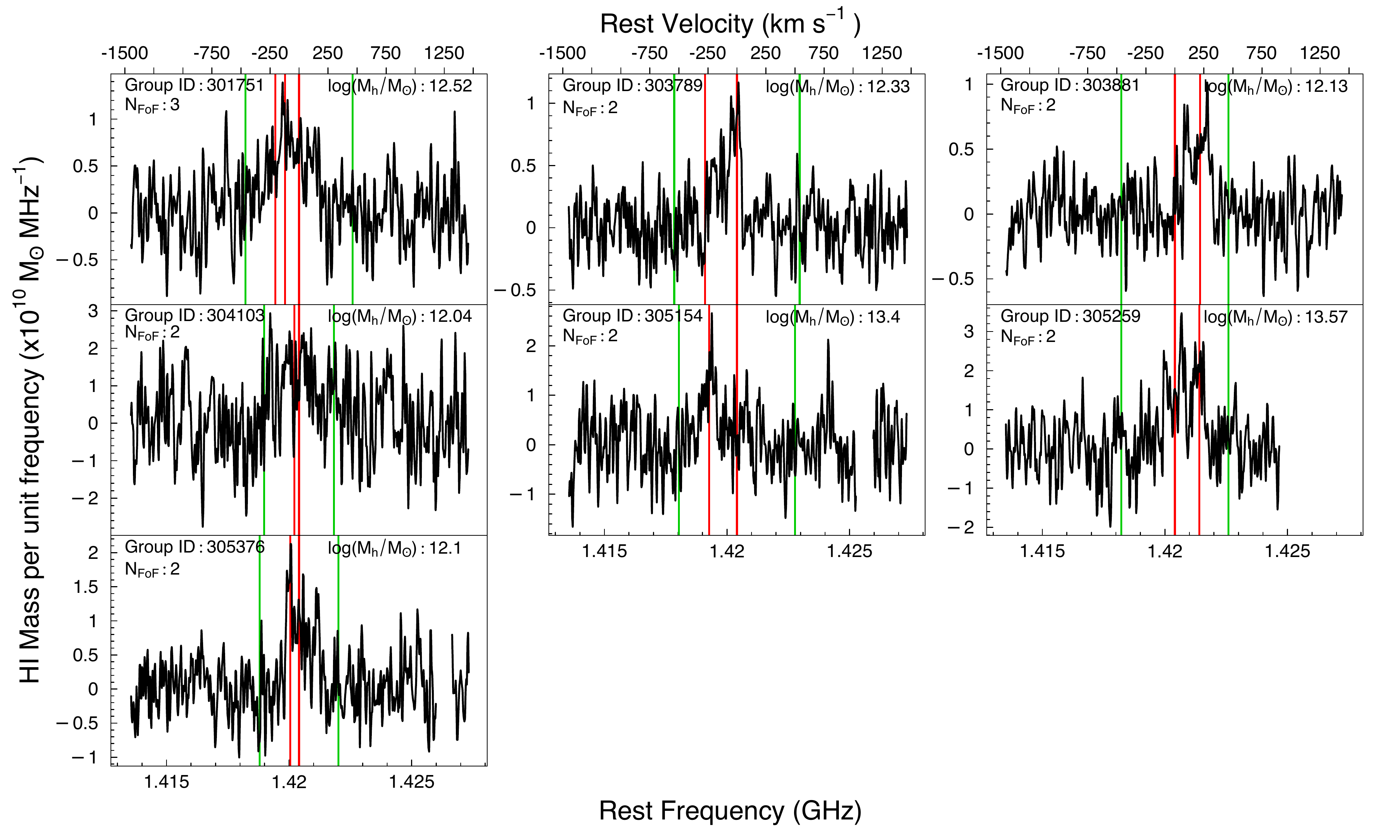}
    \caption*{Continued}
    \label{fig:det3}
\end{figure*}

\label{app:Appendix1}